# Multi-Filament Inflows Fuel Young Star Forming Galaxies

D. Christopher Martin[1], Donal O'Sullivan[1], Mateusz Matuszewski[1], Erika Hamden[1], Avishai Dekel[2], Sharon Lapiner[2], Patrick Morrissey[1], James D. Neill[1], Sebastiano Cantalupo[3], J. Xavier Prochaska[4,5], Charles Steidel[6], Ryan Trainor[7], Anna Moore[8], Daniel Ceverino[9], Joel Primack[10], Luca Rizzi[11]

[1]Cahill Center for Astrophysics, California Institute of Technology, 1216 East California Boulevard, Mail Code 278-17, Pasadena, California 91125, USA. [2]Racah Institute of Physics, The Hebrew University of Jerusalem, Israel 91904. [3]Caltech Optical Observatories, Cahill Center for Astrophysics, California Institute of Technology, 1216 East California Boulevard, Mail Code 11-17, Pasadena, California 91125, USA. [3]ETH Zurich, Institute for Astronomy, Wolfgang-Pauli-Strasse 27 8093, Zurich, Switzerland. [4]Department of Astronomy and Astrophysics, University of California, 1156 High Street, Santa Cruz, CA 95064, USA. [5]University of California Observatories, Lick Observatory, 1156 High Street, Santa Cruz. [6]Cahill Center for Astrophysics, California Institute of Technology, 1216 East California Boulevard, Mail Code 249-17, Pasadena, California 91125, USA [7]Department of Astronomy, University of California, Berkeley, 501 Campbell Hall, Berkeley, CA 94720, USA. [8]Research School of Astronomy and Astrophysics, The Australian National University, Canberra, ACT, Australia. [9]Universitat Heidelberg, Zentrum fur Astronomie, Institut fur Theoretische Astrophysik, Albert-Ueberle-Str. 2, 69120 Heidelberg, Germany. [10]Department of Physics, University of California, Santa Cruz, CA 95064, USA. [11]W. M. Keck Observatory, Waimea, HI 96743

**Theory suggests that there are two primary modes of accretion through which dark matter halos acquire the gas to form and fuel galaxies, hot and cold mode accretion. In cold mode accretion, gas streams along cosmic web filaments to the center of the halo, allowing for the efficient delivery of star-forming fuel. Recently, two QSO-illuminated HI Lyman alpha (Lyα) emitting objects were reported to have properties of cold, rotating structures[1,2]. However, the spatial and spectral resolution available was insufficient to constrain radial flows associated with connecting filaments. With the Keck Cosmic Web Imager (KCWI)[3] we now have eight times the spatial resolution, permitting the detection of these in-spiraling flows. In order to detect these inflows, we introduce a suite of models which incorporate zonal radial flows, demonstrate their performance on a numerical simulation that exhibits cold-flow accretion, and show that they are an excellent match to KCWI velocity maps of two Lyα emitters observed around high-redshift quasars. These Multi-Filament Inflow models kinematically isolate zones of radial inflow that correspond to extended filamentary emission. The derived gas flux and inflow path is sufficient to fuel the inferred central galaxy star formation rate and angular momentum. Thus, our kinematic emission maps provide strong evidence for the inflow of gas from the cosmic web building galaxies at the peak of star formation.**

Theory predicts that the inflowing gas in cold streams gains angular momentum prior to entering the halo virial radius by tidal torques from the cosmic web[4-7]. Thus, gas is delivered with significant angular momentum, producing a rotating and largely coplanar structure[4-7]. Large rotating nebulae are now being detected in Lyα emission[1,2,8] as are possible inflows[9]. At the same time the gas travels inward, necessitating a radial component to the flow. Because the gas is flowing in from discrete filaments, the spiral inflow is roughly organized in azimuthal zones associated with each filament.

A simple disk rotation model was adequate to describe the data obtained with the Palomar Cosmic Web Imager (PCWI) in prior work. [1,2] The recently commissioned Keck Cosmic Web Imager (KCWI) [3] provides a factor of eight spatial resolution gain (in number of pixels), and a factor of 10 increase in sensitivity, which allows precision velocities to be measured at this much finer spatial scale. This new capability for the first time allows us to probe higher-order components in the velocity structures of newly forming galaxies.

We require a quantitative framework to detect and characterize these higher spatial resolution components. We first describe this framework and how we have verified its efficacy by application to a mock data cube from a simulated galaxy exhibiting cold filamentary inflow. In a hierarchical decomposition of the velocity structure in a newly forming galaxy, one might expect to find at least three levels: simple rotation, radially and azimuthally varying components corresponding to the influence of inflowing filaments, and higher order effects corresponding to complex gas dynamics on smaller scales (see Fig. 20 reference [4]).

We employ a series of four increasingly complex models: (1) simple Keplerian rotation in an NFW halo, (2) rotation plus a linearly varying radial component, (3) rotation with a radial component that varies both radially and azimuthally and (4) rotation with both radial and azimuthal components both being azimuthally modulated. We test these models on a simulated protogalaxy, and then apply them to two Lyα emitting objects observed with KCWI. We display the simulated galaxy side-by-side with the two observed objects in Fig. 2 and in the model fitting Figs. 3-4, in order to compare how the addition of MFI modes improves the fit and isolates radial inflow zones in the model and in the two observed objects.

Model 1 takes the dark matter halo mass, concentration, disk inclination and position angle as free parameters. Model 2 adds a radial velocity component, proportional to radius. Model 3 allows this radial component to vary with azimuth, with an azimuthal modal decomposition of the radial velocity profile. The azimuthal variation has three modes. Model 3(a) has one cycle per revolution, with independently determined phase and amplitude. A spiral phase variation with radius can also be added. Model 3(b) is the sum of a one and two cycle component, each providing two free parameters (plus the single spiral slope). Model 3(c) adds a three-cycle component to model 3(b). In model 4 we allow the azimuthal profile to deviate from the NFW rotation model with a correction sharing the same spatial profile as the radial component.

These models are motivated by analysis of a simulated protogalaxy VELA07 [4,10,11]. Examination of the radial velocity profiles in the simulated protogalaxy shows azimuthally coherent zones apparently driven by the cold inflow filaments (Fig. 1). Radial flows show both inflow and outflow variations, with inflow dominating over the full extent of the object but outflow regions on opposite sides where gas overshoots prior to circularizing. Large scale features show a rough azimuthal modulation. On average, the radial velocity increases linearly with radius (see Supplementary Fig. 1).

Three data-cubes were generated with orthogonal viewing directions, and mean intensity-weighted velocity and velocity dispersion maps generated using a velocity window that isolates the emitting region. The resulting simulated maps are given in Fig. 2a for one line of sight.

For each perspective, we fit the MFI models to the simulation 2D velocity profile. Derived physical parameters (Supplementary Table 4) are consistent with those of the simulation. For Model 3a(i), we derive a halo mass of $\log M_h = 11.9^{+0.2}_{-0.1}$, (simulation halo mass: $\log M_h = 11.95$) and infer a mass-weighted radial flux within 50 kpc of $-24^{+6}_{-11}\ M_{\odot} yr^{-1}$ (actual flux: $-24\ M_{\odot} yr^{-1}$). The derived radial inflow direction corresponds to one of the major inflow filaments in the simulated protogalaxy. A prominent feature of the velocity maps (Fig. 3) is a distortion of the classical disk "spider diagram" such that the velocity contours run orthogonally to their baseline directions wherever radial flow is strong.

We find that Model 1 is an adequate fit to the simulated galaxy velocity field, and that each successively more complex model invoking radial flow provides a better fit (Fig. 3, Fig. 4a, Supplementary Figs. 2-3, Supplementary Table 4). This is quantified using the Akaike Information Criterion (AIC) [12], which appropriately penalizes free parameters, and is minimum for the best model. There is a dramatic reduction in $\chi^2$ and AIC moving from Model 2 to Model 3a(i) for each viewing direction. Although there are three filaments in the VELA07 simulation, the one in the lower left corner of Fig. 1a dominates because of its large column density and width. This filament is correctly detected by the MFI formalism and appears in Supplementary Fig. 2. The halo mass and mass inflow rate are also correctly derived. Similar results are obtained with two other simulated galaxies. We conclude that the MFI model can detect filamentary radial inflow in the gas velocity field of a forming galaxy. We provide a modified set of criteria for identifying a protogalaxy exhibiting MFI, extending our protogalactic disk criteria[2].

We now turn to the KCWI observations. We observed two QSO fields with KCWI; UM287 and QSO B1009+2956. UM287 is a binary QSO field with clear and extended filamentary emission[13] in which observations with PCWI revealed an extended, rotating protodisk [1]. QSO B1009+2956 (CSO38) is a field in which extended emission was initially detected via narrowband imaging[14,15] and subsequently followed up with PCWI. The Lyα emitters are extended blobs 80 pkpc and >120 pkpc in size for CSO38 and UM287 respectively, each showing a pronounced velocity shear and separated from the QSO by a projected 50-120 pkpc.

The emitting region near CSO38 (CSO38B) shows a central concentration as well as three filamentary extensions (Fig. 2, Supplementary Fig. 4), in two cases with low velocity dispersion, consistent with inflowing filaments[1,2,16,17]. The mean velocity displays a velocity shear of ~350 km/s coinciding with the central concentration, and the object meets all the criteria set out for a rotating structure with MFI components (Methods, Supplementary Table 2). There are continuum objects two of which are galaxies near the redshift of the nebula.

Fig. 2 shows the intensity, velocity, and velocity dispersion maps for UM287. There is a large, lower level emission region and a bright extended region somewhat off-center from the overall emission. There is velocity shear over the larger region and over the bright central region. The brightest emission comes from source "C", which also has a very red velocity. There are some faint continuum sources but no continuum counterpart to the entire nebula. The Lyα emission from these sources may be enhanced by the presence of star formation. The observed properties of this object also compare favorably with the MFI criteria (Supplementary Table 2).

For CSO38B we find that Model 1 is an adequate fit to the data, and that each successively more complex model invoking radial flow provides a better fit, with the exception of Model 3c(i) (Fig. 3, Supplementary Fig. 2 and Supplementary Tables 5 and 7). There is a dramatic reduction in $\chi^2$ and AIC moving from Model 2 to Model 3a(i). The formal probability that Model 2 is a better representation of the data than 3a(i) is $< 10^{-23}$, and that Model 1 is a better representation is $< 10^{-32}$. For Model 3a(i) we derive a halo mass of $\log M_h = 11.25^{+0.11}_{-0.16}$, a mode-1 velocity amplitude of $v_{r1} = -318^{+38}_{-27}$ (an 8σ detection of the MFI radial mode), and a mass-weighted radial flux of $-39^{+6}_{-4}\ M_{\odot} yr^{-1}$ (a 6.5σ detection of radial inflow). The direction of radial inflow is along one of the extensions (Fig. 4; Supplementary Fig. 5). Halo concentration is not well constrained, and during fitting it is limited to the range $0 < \log c < 1$. The galaxy at the kinematic and intensity center has a measured star formation rate of 17 $M_{\odot} yr^{-1}$. The gas inflow has the direction and magnitude to fuel this ongoing star formation and supply baryonic angular momentum in the same direction as the rotating component. Note that a model with only radial flow is excluded, since the rotating component parameterized by the halo mass is required for a good fit. Other alternative models are discussed in Methods.

For UM287 we see a successive reduction in chi-square and AIC for each model. Model 1 gives consistent results with the PCWI data and modelling: best fit parameters for KCWI are halo mass $\log M_h = 12.7 \pm 0.05$, inclination i=65°, baryon mass $\log M_b = 11.2$, and for PCWI are $\log M_h = 13.1 \pm 0.6$, $\mathrm{i} = 70°$, and $\log M_b = 11.2$. The reduction going from model 2 to 3 is not as dramatic as for CSO38B and the VELA07 simulation. However, there is a pronounced fit improvement adding the third mode to Model 3 (Model 3c(i), AIC=432, and Model 4 AIC=351). The formal probability that Model 3b(ii) is a better representation than 3c(i) is <$3 \times 10^{-17}$, and that Model 1 is a better representation is <$5 \times 10^{-73}$. Even the notional intensity model shows some agreement with the intensity map, suggesting that the morphology of the brightest emission is produced by filamentary inflow. The final velocity residual is comparable to CSO38B ($\sigma_v = 72\ km/s$). The mode-3 component has a velocity amplitude $v_{r3} = 1050^{+61}_{-133}$, an 8σ detection of this component. Inferred physical parameters for Model 4 suggest a higher mass halo, with $\log M_h = 12.69^{+0.04}_{-0.16}$, consistent with the larger size of the object. Object C coincides with a positive radial flow consistent with an outflow produced by a starburst or AGN. The mass-weighted radial flux of $-26^{+4}_{-7}\ M_{\odot} yr^{-1}$, a 6.5σ detection of inflow, is comparable to the star formation rate inferred for source D and E [1], and the radial inflow direction is along the major filament[1,13] (Fig.4, Supplementary Fig. 6, Supplementary Tables 6-7). Again, the inflow supplies baryonic angular momentum consistent with that of the rotating component.

The excellent fit of MFI models with the VELA07 simulation demonstrates that cold spiral inflow can be detected in kinematic Lyα emission line maps. As we discuss in Methods, radiative transfer effects, galactic winds, tidal interactions in the gas-galaxy systems, and the selection of QSO-illuminated objects do not significantly affect the conclusions, and that the cold inflow scenario incorporates both smooth accretion and clumpy satellite accretion. One azimuthal mode provides a good fit to the simulation and CSO38B, consistent with a single filament dominating the inflow. The presence of three modes in UM287 is consistent with the observation that 2-3 prominent filaments tend to appear in cold accretion inflow models[18] and may be a result of higher mass, larger size (and correspondingly improved resolution of the complex velocity field), different geometry, or other factors. Theory suggests that this halo mass is a regime in which cold accretion flows penetrate a hot virialized halo, which could help explain the increased complexity of the flow [19]. The MFI components and radial influx are detected with high significance even when we assume the turbulent velocity field in the objects determines the rms velocity error and AIC. The data has high signal-to-noise ratio and Poisson noise does not impact the detection significance of the MFI components. It is the much higher spatial resolution of the KCWI data that has allowed us to detect the MFI components. In both objects the detected inflow connects an extended filament with a flow direction and magnitude consistent with the star formation rate in the central galaxy and the angular momentum of the rotating gas component. Thus these observations and MFI modelling using detailed emission-line velocity maps demonstrate that cold inspiral accretion is a dominant mode of gas delivery for star forming galaxies at high redshift.

It is important to note that we are not attempting here a thorough quantitative comparison of simulations and observations. The one simulation addressed here, as an example, serves as an indicator of likely qualitative consistency between the parametric model as applied to the observations and robust features extracted from cosmological simulations. It is an important step in interpreting the model/observational results in terms of a picture of galaxy formation within the cosmic web that emerges from simulations and theory.

**Author Information** Reprints and permissions information is available at www.nature.com/reprints. The authors declare no competing financial interests. Correspondence and requests for materials should be addressed to D.C.M. (martinc@caltech.edu).

**Author Contributions.** D.C.M. is the Principal Investigator of KCWI, performed the analysis of the simulated galaxy, data and MFI models, and was principal author on the paper. D.O. and D.C.M led the observations of UM287 and CSO 38. D.O., M.M. reduced the data. D. O., M. M., and E. H. contributed to the paper writing. P.M., M.M., D.C.M., D.N., D.O., and A.M. designed, constructed, and operated KCWI. D.N., M.M. and D.C.M. developed the KCWI data pipeline and produced the final data cubes.  D. C. developed the VELA simulations. A.D. and S. L. provided the simulated galaxy VELA07 and contributed to the paper. D.C. and J.P. contributed the cosmological simulation. C.S, R.T., S.C., and J.X.P. contributed to the development of KCWI, Keck data for the two protogalaxies, and to the editing of the paper. L. R. made major contributions to KCWI commissioning and participated in the observations of CSO 38.

**Acknowledgements** This work was supported by the National Science Foundation, the W. M Keck Observatory and the California Institute of Technology.  The VELA simulations were performed at NASA Advanced Supercomputing (NAS) at NASA Ames Research Center. D.C. is funded by the ERC Advanced Grant, STARLIGHT: Formation of the First Stars (project number 339177).

## Figure Legends

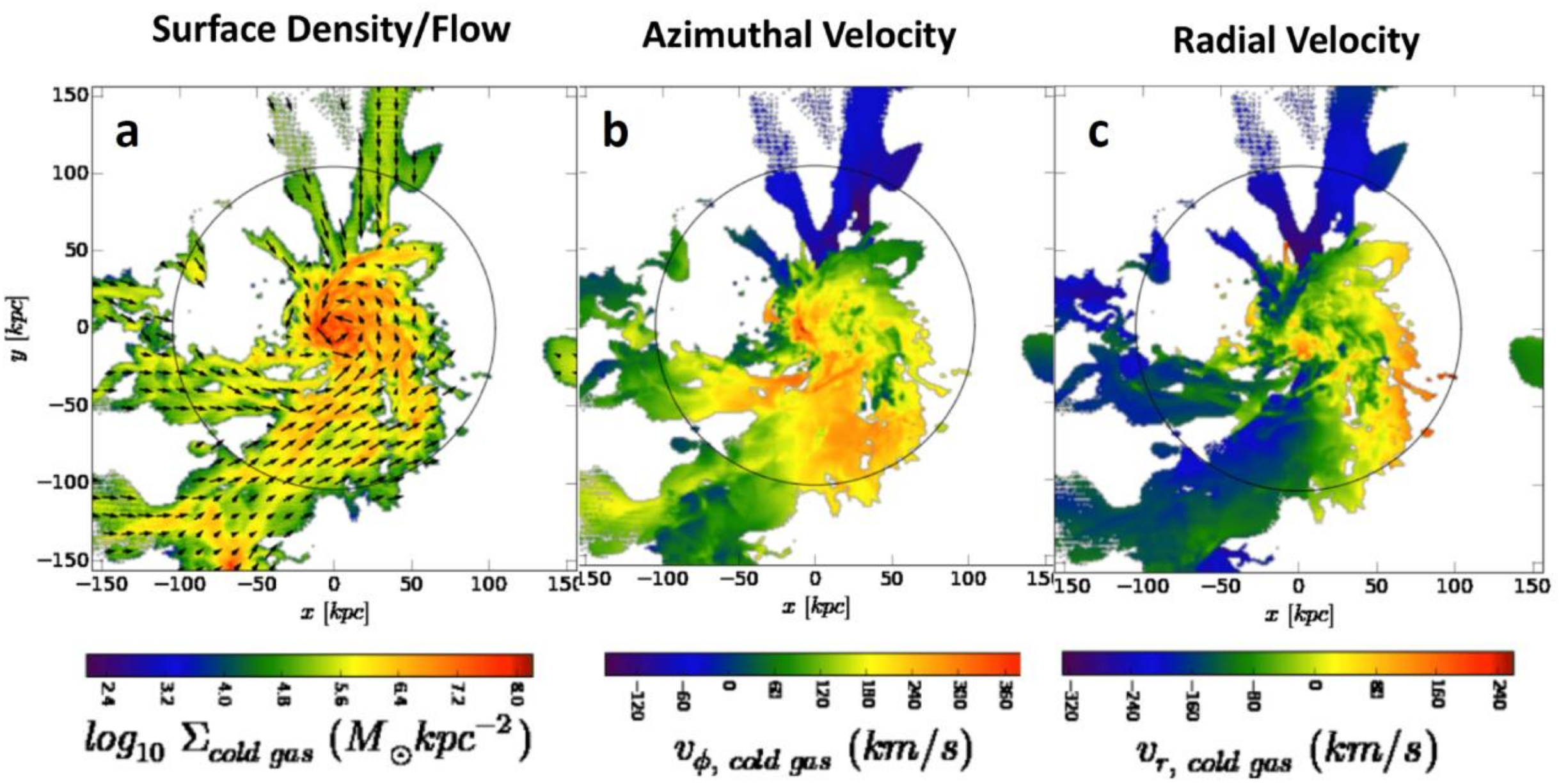

**Fig. 1 – Surface density and velocity maps of VELA07 simulation in face-on projection.** A. Column density, given on a logarithmic scale in $M_{\odot}\ kpc^{-2}$. Arrows give local gas velocity magnitude and direction in the plane of the image. Circles in this and other panels are one virial radius. b. Color value gives azimuthal gas velocity, with positive values corresponding to counter-clockwise. c. Color value gives radial gas velocity, with negative values corresponding to inflow. All maps are 300 pkpc by 300 pkpc.

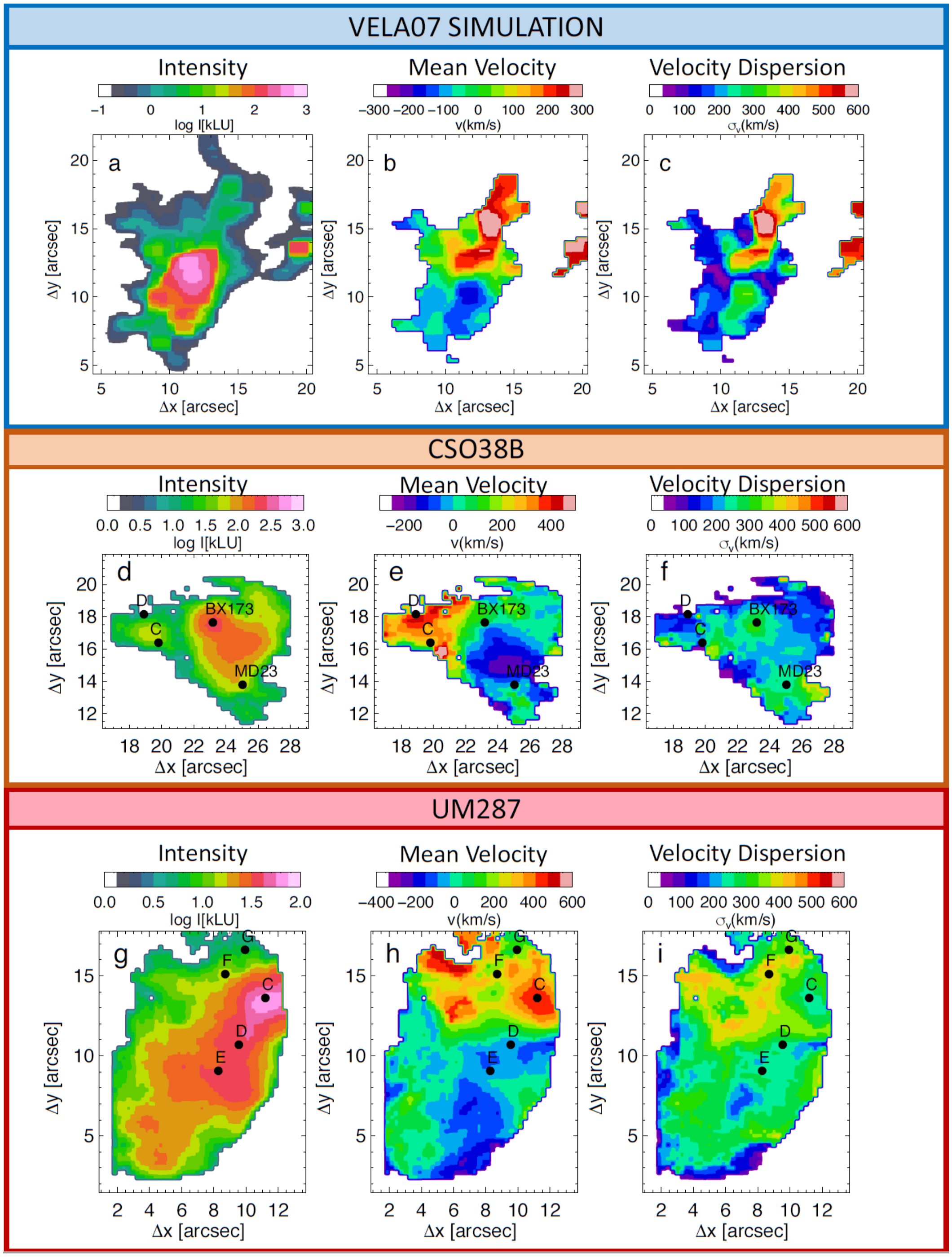

**Fig. 2 – Narrow-band image, mean velocity map and velocity dispersion maps simulation and data.** a-c. VELA07 simulation. a. Simulated narrow-band Lyα intensity map for the VELA07 simulation (in kLU). b. Mean (intensity weighted) velocity, VELA07 simulation. c. Velocity dispersion (intensity weighted), VELA07 simulation. d-f. CSO38B. d. Narrow-band Lyα intensity map for CSO38B (in kLU). b. Mean (intensity weighted) velocity, CSO38B. c. Velocity dispersion (intensity weighted), CSO38B. Continuum objects in the field also shown in this and panels g-h. See *Methods* for discussion of continuum object properties. g-i. UM287. g. Narrow-band Lyα intensity map for UM287 (in kLU). b. Mean (intensity weighted) velocity, UM287. c. Velocity dispersion (intensity weighted), UM287.

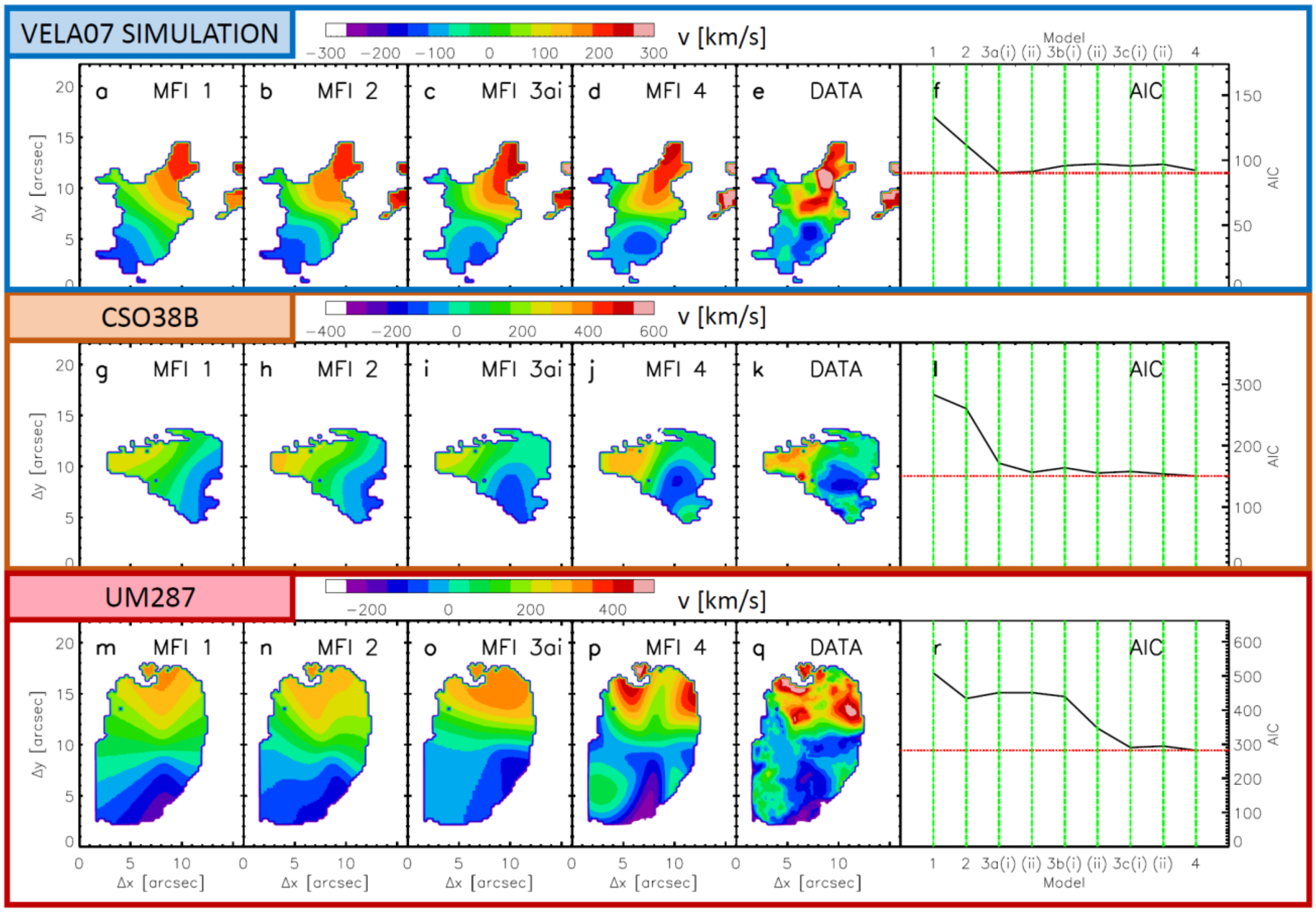

**Fig. 3 – Velocity maps simulation and data compared to Multi-Filament Inflow (MFI) models.** a-f. VELA07 simulation. a. MFI Model 1 velocity map. b. MFI Model 2. c. MFI Model 3a(i). d. MFI Model 4 (Models 3a(ii),3b, and 3c not shown). e. Simulated VELA07 velocity map (as in Fig. 2b). f. Akaike Information Criterion (AIC) vs. MFI model. Red dashed horizontal line shows minimum AIC (Model 3a(i)). g-l. CSO38B. g. MFI Model 1 velocity map. h. MFI Model 2. i. MFI Model 3a(i). j. MFI Model 4 (Models 3a(ii),3b, and 3c not shown). k. CSO38B velocity map (as in Fig. 2e). l. Akaike Information Criterion (AIC) vs. MFI model. Red dashed horizontal line shows minimum AIC (Model 4). m-r. UM287. m. MFI Model 1 velocity map. n. MFI Model 2. o. MFI Model 3a(i). p. MFI Model 4. q. UM287 velocity map (as in Fig. 2h). r. Akaike Information Criterion (AIC) vs. MFI model. Red dashed horizontal line shows minimum AIC (Model 4).

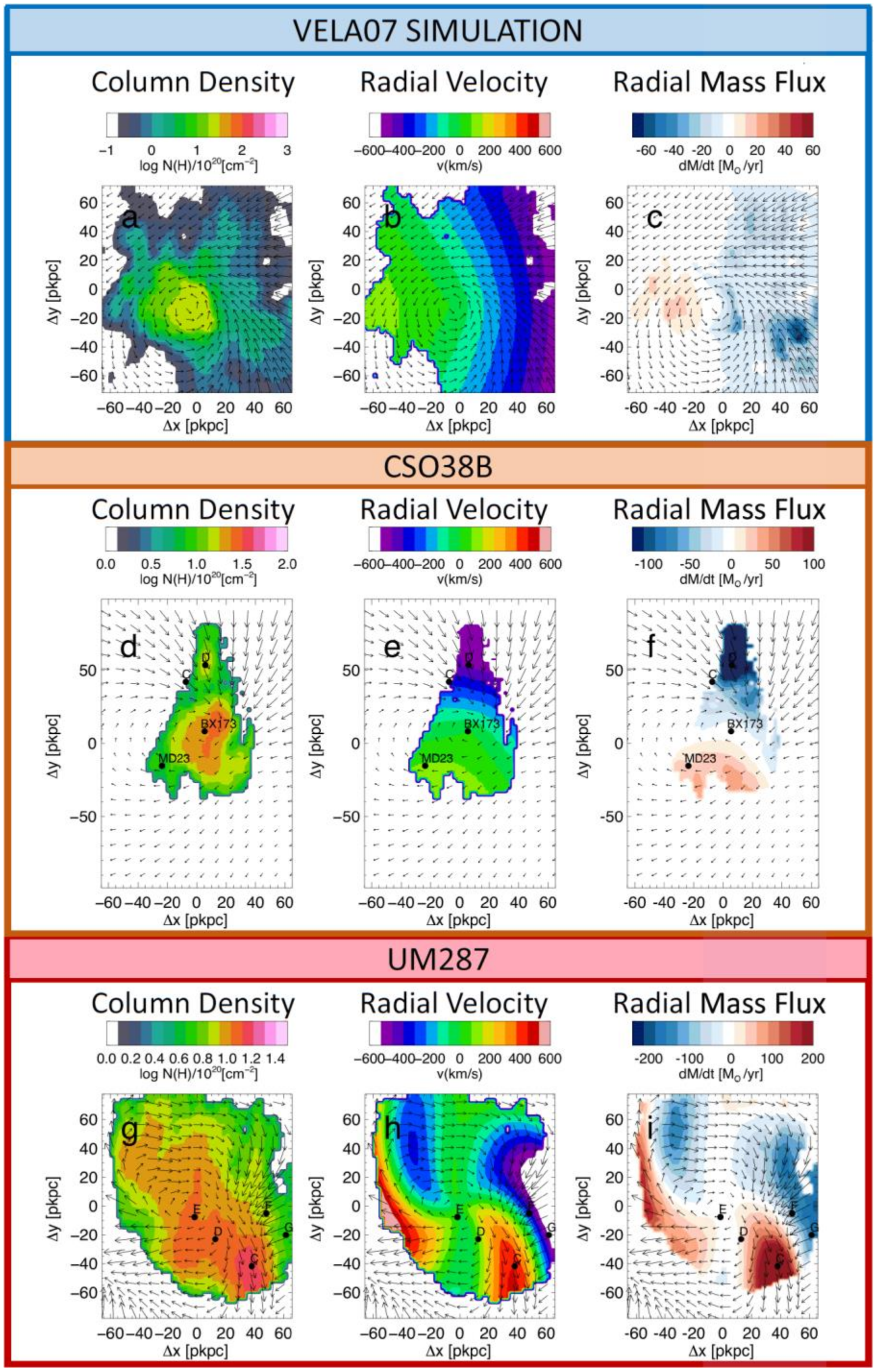

**Fig. 4 – Face-on reconstruction of column density, radial velocity, and radial mass flux for simulated and observed objects.** All plots show the inferred total velocity field with arrows. The VELA07 simulation and CSO38B use Model 3a(i), and UM287 uses Model 4. a. VELA07 simulation column density for Model 3a(i), b. VELA07 simulation radial velocity component from Model 3a(i), c. VELA07 simulation radial mass flux. d. CSO38B column density, e. CSO38B radial velocity component from Model 3a(i), f. CSO38B radial mass flux. g. UM287 nebula column density, e. UM287 nebula radial velocity component, f. UM287 radial mass flux. In all cases the mass flux amplitude plotted is $2\pi d\dot{M}/d\phi$ (cf. *Methods*). The total mass flux through radius $r$ is the average of the plotted flux on the circle of radius $r$. Continuum objects remapped to face-on position and superposed.

## Methods

### Observations

KCWI is an integral field spectrograph (IFS) designed for observing extremely low surface brightness objects[3] mounted at the Nasmyth focus of Keck II. In a 1 hour observation, with the Medium slicer and BM grating, the 5σ limiting emission line flux is 6 x $10^{-18}$ erg $cm^{-2}$ $s^{-1}$ $arcsec^{-2}$ in a 1″x1″ spatial bin, assuming a 4Å line width.

A summary of the observations, along with target information, is presented in Supplementary Table 1. KCWI observations of CSO 38 were made with the large image slicer and medium resolution grating ($\Delta\theta \sim 0.7''$, $\Delta\lambda \sim 2.5$Å, $FOV = 16'' \times 20''$) to a limiting sensitivity of 3 x $10^{-18}$ erg $cm^{-2}$ $s^{-1}$ $arcsec^{-2}$ (1″x1″) in a 4Å line. KCWI observation of UM287 were made with the medium image slicer and medium-resolution grating ($\Delta\theta \sim 0.6''$, $\Delta\lambda \sim 1$Å, $FOV = 33'' \times 20''$) to a limiting sensitivity of 3 x $10^{-18}$ erg $cm^{-2}$ $s^{-1}$ $arcsec^{-2}$ (1″x1″) in a 4Å line. We discuss the methodology and details of our observations and data analysis extensively in previous work[1,2,16,17]. The KCWI data reduction pipeline produces 3D flux cubes (α, δ, λ) which we then adaptively smooth to optimally extract emission on different spatial scales.

For the UM287 observation, separate sky images have been interleaved between the science target observations and spatially smoothed to reduce the sky noise introduced by subtraction. The nod-and-shuffle mode provides exquisite sky subtraction, but we have found for these brighter Lyα objects (typical intensities 30-100 kLU) that the throughput of KCWI is high enough and the sky background at Keck is low enough that sky subtraction of <0.5% is sufficient for excellent detections (corresponding to an additional error of ~1 kLU). A three-point dithering pattern of -0.5, 0, and +0.5 slice-widths perpendicular to the slices is used to obtain better PSF sampling in the coadded cube.

### Data Analysis

The KCWI pipeline is open source and can be downloaded from https://github.com/kcwidev/kderp. It is written in IDL and consists of 8 stages. Raw images are bias and overscan subtracted, gain-corrected, trimmed and cosmic ray removed. Dark subtraction and scatter light removal is performed. Calibration images taken with the same settings are used to define the geometric transformations to map each pixel in the 2D image to slice, position, and wavelength position. Flat field and illumination corrections are performed. Sky subtractions can also be performed, either via nod-and-shuffle, a separate sky field, or using the image itself to generate a model of the sky, which is then subtracted. The user can select which sky subtraction method to use. Cubes are generated for intensity, variance, and mask images. Differential atmospheric refraction is corrected for base on airmass, slicer orientation, and wavelength. If a standard star has been observed, flux calibration will be performed as a final step.

The KCWI data reduction pipeline produces 3D flux cubes ($\alpha$, $\delta$, $\lambda$) for each image. In the case of multiple images per field, these data cubes (and their corresponding variance images) need to be rectified and stacked. With bright quasars, the stacking process is relatively straightforward, with the brightest object lined up for all images. Images and stacks are compared to other observations of the field to verify the presence of any background continuum objects and double-check the WCS. Images are re-binned to create square pixels.

Because the QSOs in both the QSO1009 and UM287 fields was well separated from the Ly$\alpha$ emission, no QSO subtraction was required.

**Adaptive Kernel Smoothing**

After stacking and sky subtraction, the final data cube for each target is run through an adaptive smoothing process[1,2,16,17]. Adaptive Kernel Smoothing (AKS) is a process for optimally extracting signal that exists at different spatial scales within an image. This is well suited to CGM/IGM studies, as we expect both large scale structure and small-scale clumping. In two dimensions, AKS begins by smoothing data with the kernel (e.g. Gaussian) set at a user-specified minimum scale. Pixels which reach a user-specified signal-to-noise threshold are ‘detected’ at this scale. Their smoothed values are subtracted from the ‘working’ image and added to a ‘detection’ image, and the pixel is masked from future detections so that it can only contribute to further layers via smoothed residuals accumulating on larger scales. The kernel grows in size each iteration until either all of the pixels are masked or a maximum kernel size has been reached – at which point the smoothed background is added to the ‘detection’ image. The scale at each location in the image is smoothed at its own natural scale. In this way, the scale at each location is the size of the kernel (centered on that pixel) that one would need to sum under to reach the desired signal-to-noise ratio. Thus, an area with bright signal would be minimally smoothed, whereas background areas with little or no signal would be smoothed at larger scales. In three dimensions, this 2D process occurs at each wavelength bin, and a third layer of iteration is added in which the cube is wavelength-binned or smoothed. Starting at the smallest wavelength scale, the 2D process occurs at each layer of the cube. The size of the wavelength kernel is increased and the 2D process repeated at each layer again, until either all of the voxels are detected or a maximum wavelength scale has been reached. For the data in this paper, a 2D Gaussian kernel was used for spatial smoothing and a 1D boxcar kernel for wavelength.

For CSO38B, the signal-to-noise ratio threshold was 3.6. For UM287, the threshold was slightly higher, 5.1. For both objects, the smoothing window ranges from 1.5 arsec to 5.8 arcsec, with almost all of the signal in the region fitted smoothed with a kernel size of 1.5-2.6 arcsec. We note that the effective sampling of the CSO38 data is similar to that of UM287 since it was obtained in two orthogonal orientations. The minimum smoothing scale is set to 1.5", making the data sets essentially equivalent. Also, for the best fit models, the variations in the velocity map are small at <1.3" scales, so the sampling difference is not significant.

**Velocity, Velocity Error, and Dispersion Maps**

Narrow band flux, and intensity weighted velocity and velocity dispersion maps are created by summing over a velocity window roughly 1000 km/s around the intensity-weighted mean velocity of the Ly$\alpha$ nebula. This procedure is designed to ensure that in each pixel a single gas component dominates so that the velocity mean is not perturbed by multiple superposed components. Given the high signal-to-noise ratio, the mean velocity error is typically less than 10-15 km/s. The outer regions of the images are trimmed to exclude regions with intensity less than 7 kLU (CSO38B) or 4 kLU (UM287), and velocity error greater than 50 km/s. The black regions in all the maps (eg. Fig. 2, Fig. 4, etc.) are below either these thresholds or the smoothing threshold. Because both objects are quite bright, and fall off in intensity quite rapidly at their boundaries, the area fitted and the resulting best fit parameters and $\chi^2$ are only weakly dependent on this trim threshold.

We have examined the effect of a higher SNR threshold on both objects. We doubled the SNR threshold from 7 kLU to 14 kLU for CSO38B. We show in Supplementary Fig. 9a that the dramatic drop in AIC from Model 2 to Model 3a(i) is unchanged. For UM287, we perform the same experiment, doubling the SNR threshold from 4 kLU to 8 kLU. Supplementary Fig. 9b shows that the AIC drop upon reaching Model 3c(i) is essentially the same. The conclusions of the paper are thus not sensitive to small changes in the SNR threshold and are not unduly influenced by the lowest SNR data.

Velocity error maps are generated from signal-to-noise ratio cubes that are an output of the smoothing algorithm. Examples are shown in Supplementary Figure 7, along with a more detailed discussion on sources of errors in the Supplementary Notes. In brief, we use the rms velocity residual as a conservative estimate, as statistical error and sky subtraction error are found to be negligible by comparison.

**Protogalaxy Numerical Simulation**

Our cosmological simulations utilize the hydro-gravitational code ART[20 21,22] which uses adaptive mesh refinement (AMR) to follow the Eulerian gas dynamics. The code implements sub-grid models of the key physical processes relevant for galaxy formation. These include gas cooling by atomic hydrogen and helium, molecular hydrogen and metals, and photo-ionization heating by a UV background with partial self-shielding. Star formation is stochastic in cells with gas temperature $T < 10^4$ K and densities $n_H > 1\ cm^{-3}$ , at a rate consistent with the Kennicutt-Schmidt law [23]. Stellar mass loss and metal enrichment of the ISM are included. Feedback from stellar winds and supernovae is implemented by injecting thermal energy to the neighboring gas at a constant rate. Radiative stellar feedback is implemented at a moderate level, with no significant infrared trapping [11,24] .

The cosmological model adopted in the simulation is the standard CDM model with the WMAP5 cosmological parameters ($\Omega_m = 0.27$, $\Lambda= 0.73$, $\Omega_b = 0.045$, $h = 0.7$ , $\sigma_8=0.82$) [25]. Individual haloes were selected at z=1 from an N-body dark-matter-only simulation of a large cosmological box. Each halo and its environment were re-simulated at higher resolution with gas and the associated baryonic processes. The dark matter in each halo, out to a few virial radii, is typically represented by $\sim 5 \times 10^7$ particles of mass $8.3 \times 10^4 M_\odot$ each. The particles representing stars have a minimum mass of $10^3 M_\odot$. The AMR cells in the dense regions are refined to a minimum size in the range 17.5-35 pc at all times. The adaptive refinement algorithm is such that a cell is divided to 8 cells once it contains a mass in stars and dark-matter more than $2 \times 10^5 M_\odot$ , or a gas mass larger than $1.5 \times 10^6 M_\odot$. The force resolution is 1−2 grid cells of the maximum resolution. Artificial fragmentation on the cell size is prevented by introducing a pressure floor, which ensures that the Jeans scale is resolved by at least 7 cells [26].

**VELA 07 and Conversion into Observed Quantities.**

The simulated galaxy analyzed here is VELA 07 at z=2, referring to gas of T < 5 x $10^4$K. This galaxy at z=2 has a virial mass of 0.9x$10^{12}$ $M_\odot$, corresponding to a virial radius of 104 kpc, Its stellar mass is 5.7x$10^{10}$ $M_\odot$ with an effective radius of 2.8 kpc. In order to compare to more massive galaxies all quantities should be scaled up.

The VELA07 simulation snapshot consists of a list of cells with gas density, temperature, velocity, and cell size. From this list we create an observed intensity cube assuming that the gas is ionized by the nearby QSO and is emitting recombination Lyα in ionization equilibrium. The fundamental equation for the intensity/velocity cube for a given orientation (expressed in LU) is, for simulation cell $i$:

$$I(x_i, y_i, v_i) = I(x_i, y_i, v_i) + \left[\frac{\alpha f_i}{4\pi(1+z)^3}\right] n_i^2 l_i \left(\frac{l_i}{L}\right)^2$$

Where $x_i$ and $y_i$ are the projected position, $v_i$ the line-of-sight velocity, $n_i$ the gas density, $l_i$ the cell size, $L$ the pixel size in the intensity cube, and $f_i < 1$ corrects for the ionization fraction at high densities ($f_i \cong 1$ for all but the highest densities). For simplicity we assume that the gas is at a fixed temperature (T=20,000K). The intensity/velocity cube for each orientation is used to generate an intensity weighted mean velocity map. Radiative transfer effects are discussed below.

The maps shown in Fig. 2 and Fig. 4 also include the most prominent star forming regions in the simulation. Source A, at the maximum of the intensity and gas column density profile, is a galaxy with stellar mass $6.3 \times 10^{10}\ M_\odot$, star formation rate 27 $M_\odot yr^{-1}$, and mass-weighted stellar age of 1.0 Gyr. Source B, offset from center and not present in Fig. 4, is an object with stellar mass $8.5 \times 10^8\ M_\odot$, star formation rate 0.26 $M_\odot yr^{-1}$, and mass-weighted stellar age of 1.2 Gyr. Finally, object C, also offset, has stellar mass $7.3 \times 10^8\ M_\odot$, star formation rate 0.21 $M_\odot yr^{-1}$, and mass-weighted stellar age of 1.4 Gyr.

We have also analyzed two other simulated galaxies,  VELA20 and VELA21. The column density distributions are shown in Supplementary Fig. 10. The dependence of AIC on the MFI model is shown in Supplementary Fig. 11 for three different line of sight directions for all three VELA simulated galaxies. We confirm that MFI models provide a far better representation for all three simulated galaxies by finding a pronounced drop in AIC as we pass from Model 1 to Model 3 and 4 in each case and for all lines of sight.

**Radiative Transfer Effects**

We have assumed that the velocity maps correctly represent the underlying gas kinematics. As in ref. [1], we do not include any radiative transfer effects on the escape of Lya. We note that proximate QSO illumination leads to high ionization ratios and relatively low HI column densities. For CSO38B, we estimate that at a proper distance 100 pkpc the ionization rate $\Gamma = 2 \times 10^{-8} s^{-1}$, which for a maximum column density of $N_H = 10^{21} cm^{-2}$, slab thickness of 3 kpc, and gas temperature $T = 2 \times 10^4 K$ results in a neutral hydrogen column density $N_{HI} = 4 \times 10^{14} cm^{-2}$. For UM287 this becomes $N_{HI} = 10^{15} cm^{-2}$. Even lower column densities are likely since the gas temperature is predicted to be higher ($T > 10^5 K$) due to the photoionization heating. The resulting optical depths in line center are low, $\tau \sim 2 - 20$. Since the Lya is generated by recombination in the ionized gas, it will be generated with the line center reflecting the local velocity of the gas. It is well known that the velocity center of the emerging Lya line from a slab of gas at a fixed velocity is centered on that velocity [27-29].  Thus the principal question to be addressed is whether a significant column density of neutral hydrogen is present along the line of sight that has a different velocity. In the case of the MFI model, the gas is confined to a relatively thin slab, with a single velocity associated with each radial/azimuthal postion. Thus, while the line width could be increased (or even made double) depending on the optical depth, the line center should accurately reflect the local velocity by construction. In the case of the VELA simulations, we can check by estimating the 3D distribution of HI. We used CLOUDY to estimate the dependence of the local gas temperature on the gas density. We then calculate the local HI column density in each voxel. For each of three different projections we then calculate the emissivity weighted velocity and the N(HI)-weighted velocity, and compare them. They are quite close with a slope near unity and a small dispersion of $\sim$10 km/s.  Thus even in the presence of some photon diffusion the local velocity measurement should be accurate.

**Multi-Filament Inflow Protogalaxy Criteria**

We proposed a set of criteria to evaluate whether a Lyα nebula was consistent with the previously termed "Proto-Galactic Disk" [2], presented as a Supplementary Note. Here, we revise these recognizing the characteristics of Multi-Filament Inflow Protogalaxy and then evaluate the two objects in light of the revised criteria.

**Revised MFI Protogalaxy Criteria:**

1. Higher intensity located *approximately symmetrically* around 1D velocity center and relatively uniform intensity with clear intensity break at edges;

2. Near constant slope velocity gradient (1D) consistent with an NFW halo; *Deleted since the radial flow impacts the 1D velocity profile.*

3. 2D velocity and intensity distribution consistent with a disk *and multi-filament radial flow* in an NFW halo with minimal residuals;

4. Evidence for one or more filaments with low velocity gradient and possibly lower-velocity dispersion *aligned with inferred radial flow direction(s)*;

5. *Delete and replace with: Velocity gradients transverse to rotation-induced velocity shear.*

6. Star formation *near* center of disk *with radial mass flux onto galaxy consistent with star formation rate*; and

7. Kinematics consistent with radial and spiral inflow. *Deleted, redundant with 3.Replace with: Object is separated from illuminating QSO and does not appear to be part of interaction producing the QSO.*

We summarize the two objects in the context of these revised criteria in Supplementary Table 2.

**Multi-Filament Inflow Models**

Four model classes, each successively more complex, were used to fit the mean velocity maps. A summary of the models is provided in Supplementary Table 3.

**Model 1. Simple equilibrium rotation.** Simulations show a transition from a plane for the outer halo (>$0.3R_V$) to a somewhat modified plane for the inner halo (<$0.3R_V$)[4]. The emission we observe is largely within $0.5R_V$, so for simplicity we retain a single plane. This model[1,2] thus assumes that the emitting gas lies in a flattened disk that is rotating in Keplerian equilibrium an NFW dark matter halo. The halo is determined by the halo mass and concentration. In addition, the disk inclination, position angle, and velocity center are fit as well. The object center is fixed at the center of light. Allowing the object center to be fit as well does not change any conclusions. Halo concentration is not well constrained, and the fit is limited to $0<\log c<1$. Disk thickness does not enter into the kinematic model. There are five free parameters to fit.

There is an ambiguity in the disk orientation. Only the angular momentum projected in the plane of the sky can be measured. The angular momentum vector can be oriented toward the observer, or away. For pure rotation this does not matter, but when radial flow is added the sign of the flow (inflow vs. outflow) is reversed in the two cases.

**Model 2. Rotation plus radial flow.** This model adds a radial flow component to the disk rotation velocity field of Model 1. Examination of the VELA07 simulation protogalaxy shows that the radial component of the flow inside the virial radius decreases approximately linearly with radius. We therefore assume that the radial flow velocity is linearly proportional to radius. The amplitude quoted is the radial velocity at the virial radius of the halo. In all models we assume the radial flow is in the same plane as the rotation, and constant in azimuth. This adds one additional free parameter, totalling six. We also resolve the orientation ambiguity by assuming there is net radial inflow. In all cases the infered inflow directions (in Models 3 and 4) are consistent with filamentary extensions, validating this approach.

In some simulations[30] the radial velocity is constant with radius for $r > 0.3r_V$. Note that for all three galaxies there is no velocity information for $r > 0.5r_V$. We considered two additional radial velocity dependencies for Models 2-4. The first is a simple constant independent of radius, and the second is constant for $r > 0.3r_V$ and linearly varying with radius for $r \leq 0.3r_V$. In both cases these variations produced somewhat poorer fits to the VELA07 simulation, CSO38B, and UM287. Therefore for this investigation we retain the linear dependence at all radii as the baseline model. As the gas circularizes later in the evolution of the protogalaxy, the radial component should decrease or remain constant depending on the balance of gravitational and dissipational forces.

**Model 3a. Multi-Filament Inflow, 1 mode**. Examination of Fig. 1 shows that in the simulated galaxy the radial flow amplitude varies with azimuth. This is to be expected as gas is inflowing preferentially along filaments. In this family of models we allow the radial flow to be azimuthally modulated by sinusoids. For Model 3a, the azimuthal function is a single sinusoid with one cycle per rotation. This component is added to the constant radial component of Model 2. See Table 2 for the equation. For Model 3a(i), the azimuthal phase is fixed with radius. Model 3a(i) adds two additional parameters, the amplitude and the aziumuthal phase of the modulation, for a total of eight fitting parameters. For Model 3a(ii), the azimuthal phase is permitted to vary linearly with radius, creating a spiral phase, and adding one additional parameter, for a total of nine. Note that radial flow can be inward or outward. Inward radial flow along the disk plane seems natural, while outflow does not. Examination of Fig. 1 shows azimuthal zones where radial outflow is occuring, presumably from inflowing gas that does not completely circularize before overshooting with some outward component. Outflow due to winds from star formation and/or AGN activity would be expected to be perpendicular to the disk, although when the disk is inclined could appear to have a component parallel to the disk when radiative transfer affects are included. Wind outflows are not included in the model.

Roughly speaking, Model 3a provides a good representation of a system in which a single filament is providing the bulk of the accretion inflow. Model 3b allows for two, and Model 3c for three. In the later case the relative importance and azimuthal zone of each filament is determined by the free parameters, and one filament may still dominate. While the three filament configuration is often seen to be the default [4,5,18], the number of prominent streams may be mass and redshift dependent [31]. One of our objectives in this work is to test for the presence of multiple filaments by investigating the dependence of the fit quality on the number of modes.

**Model 3b. Multi-Filament Inflow, 2 modes.** To Model 3a we add a second modulation, 2 cycles per rotation. This adds for Model 3b(i) two more parameters for 10 total, and for Model 3b(ii) the number is 11.

**Model 3c. Multi-Filament Inflow, 3 modes.** To Model 3b we add a third modulation, with 3 cycles per rotation. This adds for Model 3c(i) two more parameters for 12 total, and for Model 3b(ii) the number is 13.

**Model 4. Multi-Filament Inflow, 3 modes, azimuthal velocity modulation.** For Models 1-3 we have assumed that the azimuthal velocity at a given radius is simply determined by the enclosed dark matter mass. Examination of Fig. 1 also shows azimuthal variations of the azimuthal velocity that are correlated with the azimuthal variations in the radial velocity. We allow for this by adding one more parameter, positive or negative, which when multiplied by the total radial velocity field is added to the azimuthal velocity field. This results in 14 free parameters.

**Model Fitting.** Models are fit by minimizing the total $\chi^2$ calculated from the difference between the data and velocity model map. We use a Powell minimization routine [32] to find the minimum. With many parameters we find that starting the minimization routine with different initial parameter values can lead to different local minima. We therefor “anneal” by repeating the minimization search with randomly chosen initial parameters. We note that the minimum $\chi^2$ models all have comparable model parameters and in particular similar MFI components. This is illustrated in Supplementary Fig. 8. In this case, the three minimum $\chi^2$ fits to Model 4, UM287 are shown. Iteration 15, 0, and 1 have AIC of 351, 364, and 379, yet as the figure shows have very comparible MFI components. Thus, the MFI modes are not supressed in the some of the local minima, and are in fact quite similar in model fits with similarly low $\chi^2$. Model parameter errors are determined in the usual fashion by fixing the parameter of interest and allowing the remaining parameters to vary. Derived parameter (such as mass flux) error ranges are obtained during the above error determination. For each parameter 1-sigma limits we derive the corresponding derived parameter values. Then the maximum and minimum of the ensemble over all the parameters gives the 1-sigma range of the derived parameter.

**Akaike Information Criterion.** We use the Akaike Information Criterion[12] to judge whether a model with a larger number of parameters is a better representation of the data than a model with fewer parameters and larger $\chi^2$. For small sample size it is:

$$AIC = \chi^2 + 2p + \frac{2p(p+1)}{N-p-1}$$

Here $p$ is the number of parameters and $N$ is the number of degrees of freedom. The minimum AIC is the most likely model. The probability that Model i minimizes the information loss given the minimum AIC model is $P(AIC_i) = \exp((AIC_{min} - AIC_i)/2)$. This is the probability tabulated in Table 1 and 3. In order to be conservative we calculate the $\chi^2$ with the number of degrees of freedom determined by the seeing disk and smoothing algorithm and not simply the number of pixels, and rather than using the formal velocity error, which is quite small (~10-15 km/s), we use the minimum rms residual velocity, which is typically ~70 km/s (except for the VELA07 simulation, for which it is ~50 km/s). We have also evaluated the Bayesian Information Criteria (BIC) [33], given by:

$$BIC = \chi^2 + p\log(N)$$

These are tabulated in Supplementary Table 6, and show a very similar behavior to that of the AIC. The tabulated BIC probabilities assume that the models have equal prior probability.

**Physical Parameter Derivation.** Baryonic mass, angular momentum, and radial mass flux require an estimate of the gas column density. As we have done in earlier work[1,2], we assume that the gas is fully ionized by the QSO and situated in a roughly planar disk. The disk thickness $t$ and gas clumping factor $C$ are unknowns. Given this assumption the intensity is approximately

$$I[kLU] = 250\, N_{21}^2 t_3^{-1} C$$

Where as usual at λ=4000Å,

$$1\ \mathrm{kLU}\ =\ 10^3\ \mathrm{ph}\ cm^{-2}s^{-1}sr^{-1} \approx 10^{-19} erg\ cm^{-2}s^{-1}arcsec^{-1}.$$

We assume that the disk thickness is 3 kpc ($t_3 = 1$) and clumping factor $C$=1. The radial mass flux is given by

$$2\pi \frac{d\dot{M}}{d\phi}(r,\phi) = ArN_H(r,\phi)v_r(r,\phi)$$

Detailed tabulations of the best fit parameters and derived physical parameters for each model are given in Supplementary Tables 4-6. Parameter error ranges for the adopted models are given in Supplementary Table 7.

**Detailed Maps in Observed and Face-on Frame.** We have generated a series of maps in the observed and inferred face-on frames in order to gain more insight into the implications of the MFI model fits and derived velocity fields. These are shown for the best fit models for the VELA07 simulation (Supplementary Fig. 3), CSO38B (Supplementary Fig. 5), and UM287 (Supplementary Fig. 6). These maps show azimuthal and radial velocity components, radial mass flux, inferred column density, and velocity field maps. We note in particular that in each case the direction of the principal radial inflow aligns with directions manifesting filamentary extensions. The flow direction is such that in each case gas will be brought to the central galaxy to fuel on-going star formation.

**Continuum Objects in the Field and Impact on Conclusions.** CSO38B has four continuum sources detected in or nearby the nebula (Fig. 2). We present a fully detailed description of these sources in the Supplementary Notes. Since galaxies are present in the nebula, we need to address three questions: 1. How much of the emission is produced by the halo of the galaxy itself? 2. Do outflows from supergalactic winds have any impact on the kinematics or the conclusions? And 3. Are there tidal interactions which impact the kinematics and the conclusions of this paper?

1. The surface brightness is ~10 times higher than the brightest Lyα halo emission seen around star forming galaxies at slightly higher redshift[34], when corrected for surface brightness dimming. We conclude that the Lyα emission is powered mostly by QSO illuminated fluorescence rather than by the star forming galaxy. The emission is brightest around BX173 because the gas column density is highest.

2. We argue here that a. the emission measure in the wind is insufficient to produce a significant fluorescent signal, and b. the kinetic energy flux is insufficient to produce a significant kinematic signature except very close to the galaxy, c. a kinematic signature is not observed, except for increased velocity dispersion near the galaxy.

2a. Emission measure. We can estimate the emissions measure by assuming isotropic flow into Ω steradians and an entrained cold gas mass flow rate that is approximately equal to the star formation rate[35]. We find that

$$EM[cm^{-5}] \approx 2 \times 10^{16} \left(\frac{\dot{M}_w}{10 M_\odot yr^{-1}}\right)^2 \left(\frac{b}{10kpc}\right)^{-3} \left(\frac{v_w}{500km/s}\right)^{-2} \left(\frac{\Omega}{4\pi}\right)^{-2} C$$

Here $\dot{M}_w$ is the wind cold gas mass flux, $b$ is the projected impact parameter ($> b_{min} \approx 3kpc$ the typical size of a starburst wind base), $v_w$ is the wind velocity, and *C* is the clumping factor. This is about a factor of 10,000 below the observed emission measure for *C=1*. The radial outward flow leads to a steeply falling density profile, and the density square dependence produces a very low emission measure even if the clumping factor were much greater than unity. Note that while external fluorescent illumination is probably undetectable, without an ionizing QSO nearby Lyα produced in the galaxy can encounter a large optical depth and therefor will diffuse out producing an extended halo, as is observed in star forming galaxies[34,36]. As we noted above the surface brightness of this halo is much lower than we observe.

2b. Kinetic Energy Flux. We attempt to estimate the impact of the wind kinetic energy flux on the motions of cold gas in the galaxy. It is expected that the supergalactic winds in average mass galaxies with established disks will form a bipolar outflow along the angular momentum axis of the star forming galaxy, as this is the path of least resistance. The emission nebulae produced by these supergalactic winds are typically confined within a few kpc. To be conservative we can assume that an areal fraction $f_A \approx \frac{2\pi r t}{(4\pi r^2)}$ at radius *r* of the gas sheet of thickness *t* intercepts what we will assume is an isotropic kinetic energy flux[35] of $\dot{E_w} \approx 2 \times 10^{41} \dot{M}_*$, operating for a dynamical time for that radius (since the gas is constantly being replenished by ongoing accretion). For BX173 this gives a total kinetic energy of ~$10^{57}$ ergs. Very conservatively assuming this all goes into random velocity of the cold gas, the (1D) velocity dispersion of the gas will be $\sigma_v \approx 23\ km/s$, a relatively small impact. Other than large scale shocks moving radially outward, there is no clear mechanism for galactic winds to significantly perturb the velocity field of the large cold gas reserve. We do note that the velocity dispersion is high near the galaxy. This may be due to unresolved rotation or turbulent motion, partially due to the injection of wind energy.

2c. No kinematic signature. Other than the increase in the velocity dispersion near the galaxy, there is no kinematic signature of a large-scale galactic wind. There is no large, biconical morphology with large velocity offsets. There is no detected radial outflow centered on BX173.

These conclusions hold equally for galaxy MD23. Moreover, excluding the region near MD23 from the fit does not change the conclusions of the fit. This is because the best fit Model 3a(i) does not do a good job in the region near MD23 (Fig. 3, Supplementary Fig. 2).

3. Tidal interactions between BX173 and MD23? Can the presence of MD23 approximately 40 kpc away (in projection) impact the kinematics of the gaseous nebula? Given the relative velocity of the two galaxies, the projected distance, and assuming MD23 is at the center of a dark matter subhalo with 10 times the stellar mass of MD23, we can estimate that the velocity gradient induced by tidal forces will be at most 1 km/s/kpc. The gradient observed in the central part of the nebula is ~10 km/s/kpc. Also, as we discuss below, the morphology of the nebula does not resemble that of nearby interacting galaxies with tidal features.

UM287 shows continuum sources in the brightest part of the nebula[1]. A full description of these sources is presented in our Supplementary Notes. However, the discussion above concerning halos, winds and tidal interactions applies as well to this system.

**Smooth vs. Clumpy Accretion in a Preferred Plane**

The accretion of cold gas along streams, the scenario we are addressing in this paper, includes the rapid gas accretion along a plane of satellites. Phenomena previously described that include coplanar streams[4,5] and extended rings[4], are examples that show cold streams define a preferred plane (both in the Horizon-MareNostrum RAMSES simulation [37], and in the VELA simulation used here). It has always been argued and shown in the simulations[18] that the cold streams are partly made of smooth gas and partly clumpy, including all the merging galaxies. So the stellar and gas accretion of satellites is part of the cold spiral inflow phenomenon.

**Alternative Models**

We must consider whether alternative models can explain the observed intensity and velocity maps. In particular, both systems reside in the neighborhood of a luminous QSO. We provide an additional note on the impact of selecting such objects in the Supplementary Notes. The QSO is likely a result of a complex interaction or merger, and it is conceivable that the gas nebulae were created as a byproduct of this interaction.

For CSO38, the projected distance between CSO38 and CSO38B is 120 pkpc. There is no morphological resemblance to nearby interacting systems such as NGC4038/39[38] and the projected distance is probably too large for CSO38B to be an interaction byproduct. As we discussed above the interaction of BX173 and MD23 is unlikely to affect the kinematics significantly.

For UM287 we previously considered three classes of alternative models to a cold-flow accretion scenario[1]. We review these in light of the higher resolution KCWI data without changing our previous conclusions.

Scenario one is that the nebula is created by an interaction involving the neighboring QSO (B) leading to a merging disk and tidal tails fluoresced by QSO A. The arguments against this have not changed: no Lyα emission or complex kinematics near QSO B, and no evidence of long, thin, curved tidal tails with a continuous velocity shear and fall-back gradient[38].

Scenario two holds that the bright part of the filament is part of a merging disk hosting QSO A or the tidal tail, and that the faint filament is part of the tidal tail. The complex kinematics revealed by KCWI are even less consistent with that expected for merging rotating disks although all possible scenarios cannot be evaluated. The filament does not exhibit the curvature in velocity-position space of a tidal tail[1,38]. The bright nebula is tangential to QSO A rather than centered on it, and the higher resolution of KCWI does not reveal any extended emission centered on QSO A from the interacting host galaxy forming QSO A. A tidal tail would extend radially outward and then curve tangentially in the plane of the merging disk and QSO A, and likely show a large velocity gradient and a discontinuity at the tail/disk interface. The filament is also wider than expected (~60 pkpc vs. 10-20 pkpc).

Scenario three posits a separate interacting system not responsible for QSO A or B, but this is also ruled out by the same arguments as above. We note however that QSO A could have been produced in its own host galaxy by the tidal forces and radial flows due to the influence of the nearby forming protogalaxy and associated subhalo that we are observing.

**Code and Data Availability.** KCWI pipeline code is available on the W. M. Keck Observatory website. KCWI data on CSO38 and UM287 is publically available. Data on UM287 will be available 18 months after the observation in Oct 2017. The data that support the plots within this paper and other findings of this study are available from the corresponding author upon reasonable request.

# Supplementary Information

## Multi-Filament Inflows Fuel Young Star Forming Galaxies

D. Christopher Martin[1], Donal O'Sullivan[1], Mateusz Matuszewski[1], Erika Hamden[1], Avishai Dekel[2], Sharon Lapiner[2], Patrick Morrissey[1], James D. Neill[1], Sebastiano Cantalupo[3], J. Xavier Prochaska[4,5], Charles Steidel[6], Ryan Trainor[7], Anna Moore[8], Daniel Ceverino[9], Joel Primack[10], Luca Rizzi[11]

[1]Cahill Center for Astrophysics, California Institute of Technology, 1216 East California Boulevard, Mail Code 278-17, Pasadena, California 91125, USA. [2]Racah Institute of Physics, The Hebrew University of Jerusalem, Israel 91904. [3]Caltech Optical Observatories, Cahill Center for Astrophysics, California Institute of Technology, 1216 East California Boulevard, Mail Code 11-17, Pasadena, California 91125, USA. [3]ETH Zurich, Institute for Astronomy, Wolfgang-Pauli-Strasse 27 8093, Zurich, Switzerland. [4]Department of Astronomy and Astrophysics, University of California, 1156 High Street, Santa Cruz, CA 95064, USA. [5]University of California Observatories, Lick Observatory, 1156 High Street, Santa Cruz. [6]Cahill Center for Astrophysics, California Institute of Technology, 1216 East California Boulevard, Mail Code 249-17, Pasadena, California 91125, USA [7]Department of Astronomy, University of California, Berkeley, 501 Campbell Hall, Berkeley, CA 94720, USA. [8]Research School of Astronomy and Astrophysics, The Australian National University, Canberra, ACT, Australia. [9]Universitat Heidelberg, Zentrum fur Astronomie, Institut fur Theoretische Astrophysik, Albert-Ueberle-Str. 2, 69120 Heidelberg, Germany. [10]Department of Physics, University of California, Santa Cruz, CA 95064, USA. [11]W. M. Keck Observatory, Waimea, HI 96743

## Supplementary Figures

## Supplementary Tables

## Supplementary Notes

**Supplementary Figures**

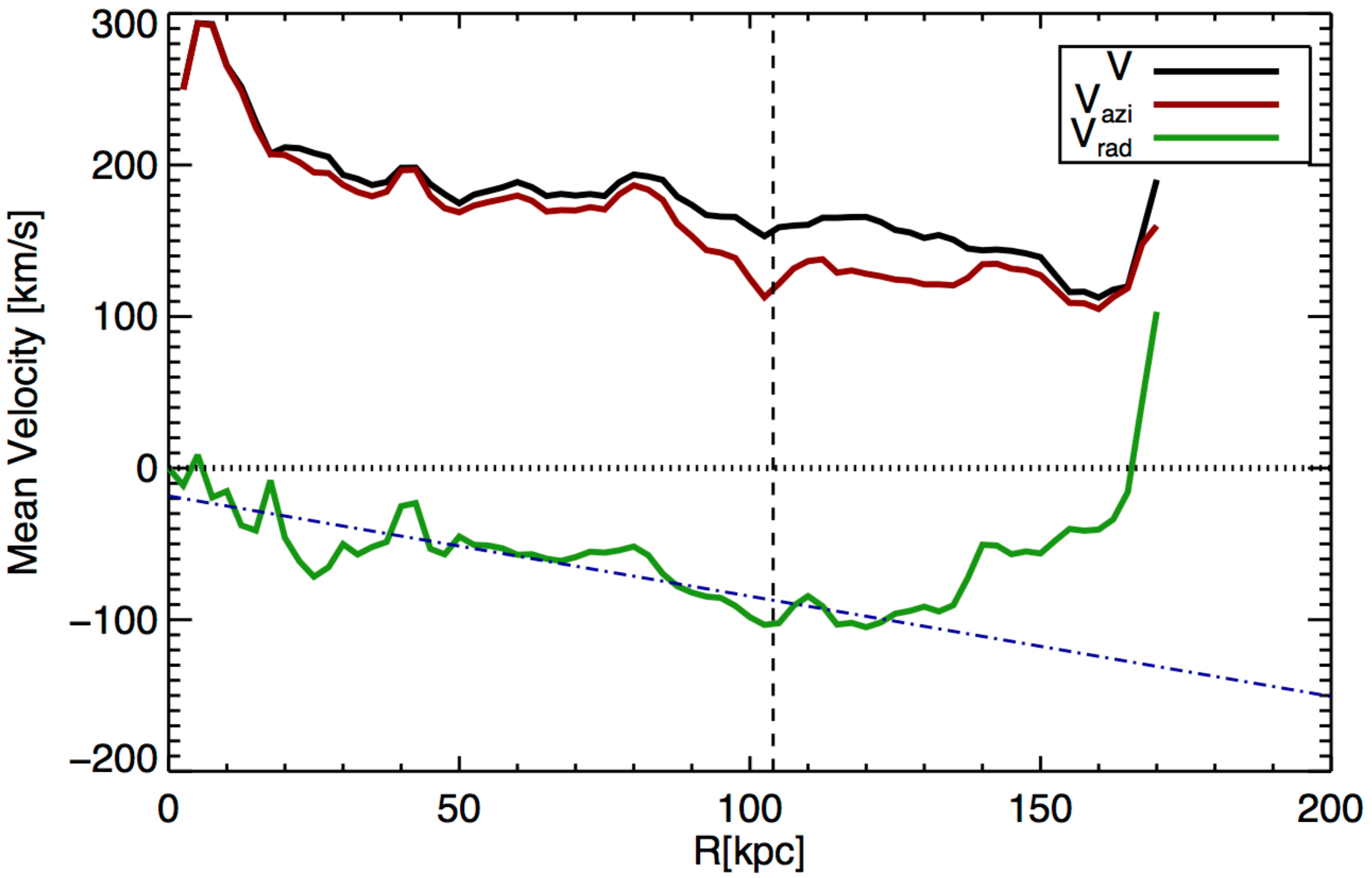

**Supplementary Figure 1** Mass-weighted mean velocity, radial velocity, and azimuthal velocity in radial bins summed over azimuth for the VELA07 simulated protogalaxy. Virial radius is shown as vertical dashed line. Radial velocity shows an approximately linear radial dependence inside the virial radius, as shown by the best fit linear function (blue dash-dot line).

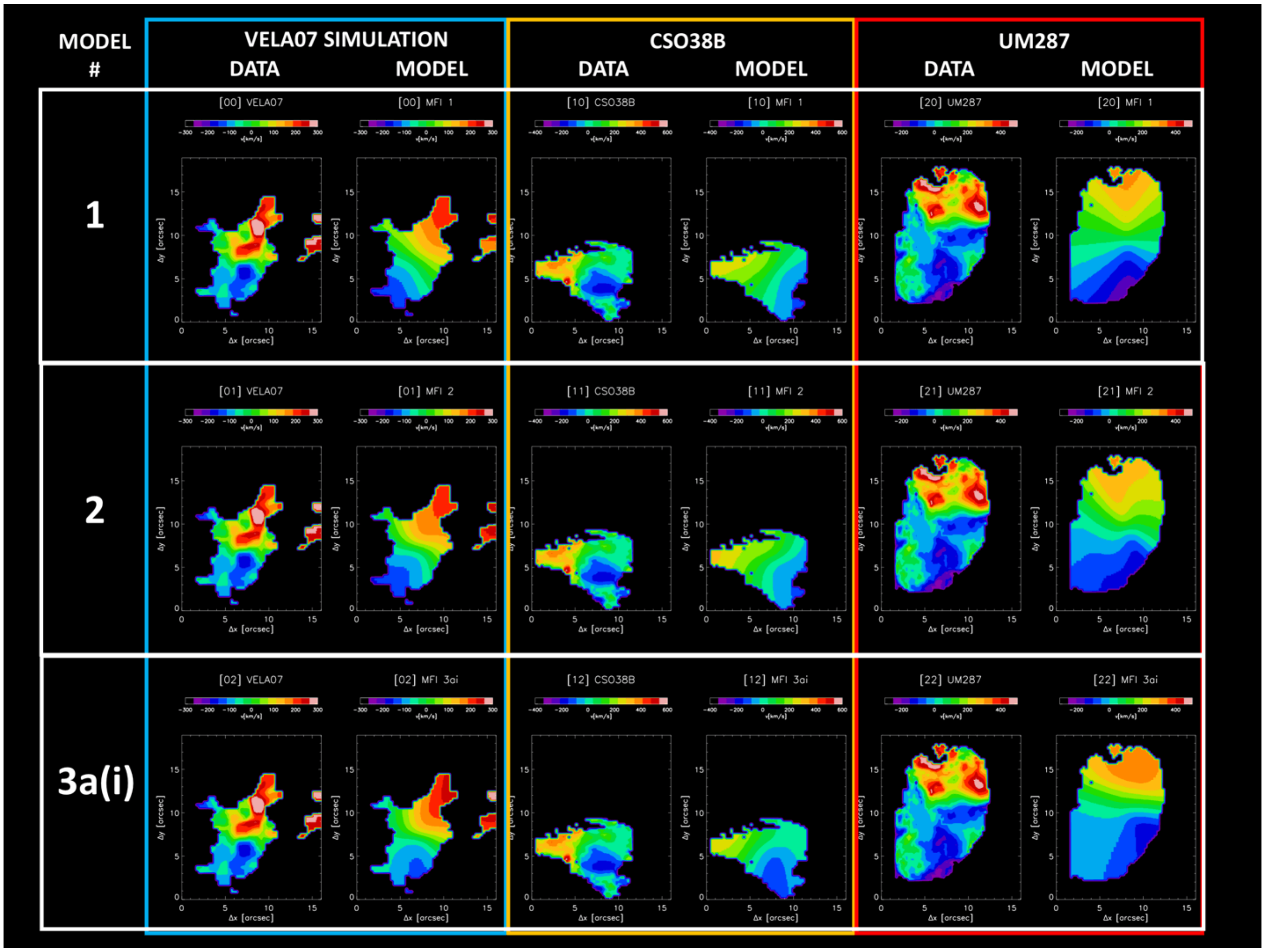

**Supplementary Figure 2a** Velocity maps for the VELA07 simulation (columns 1 and 2), CSO38B (columns 3 and 4), and UM287 (columns 5 and 6). Left-hand image in each column is data, right-hand image is model fit. a. Models 1, 2, 3a(i).

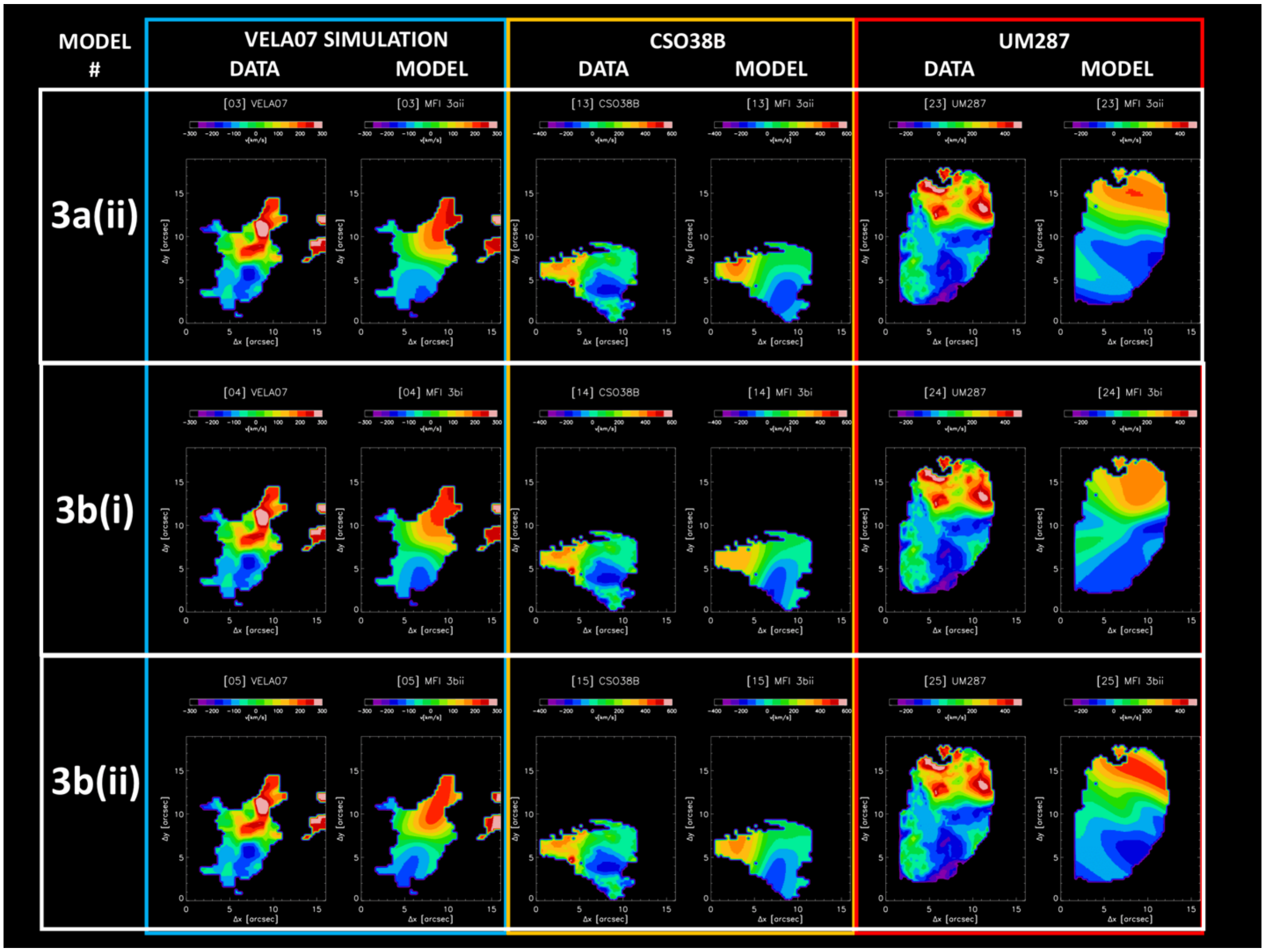

**Supplementary Figure 2b** Velocity maps for the VELA07 simulation (columns 1 and 2), CSO38B (columns 3 and 4), and UM287 (columns 5 and 6). Left-hand image in each column is data, right-hand image is model fit. Models 3a(ii), 3b(i), 3b(ii).

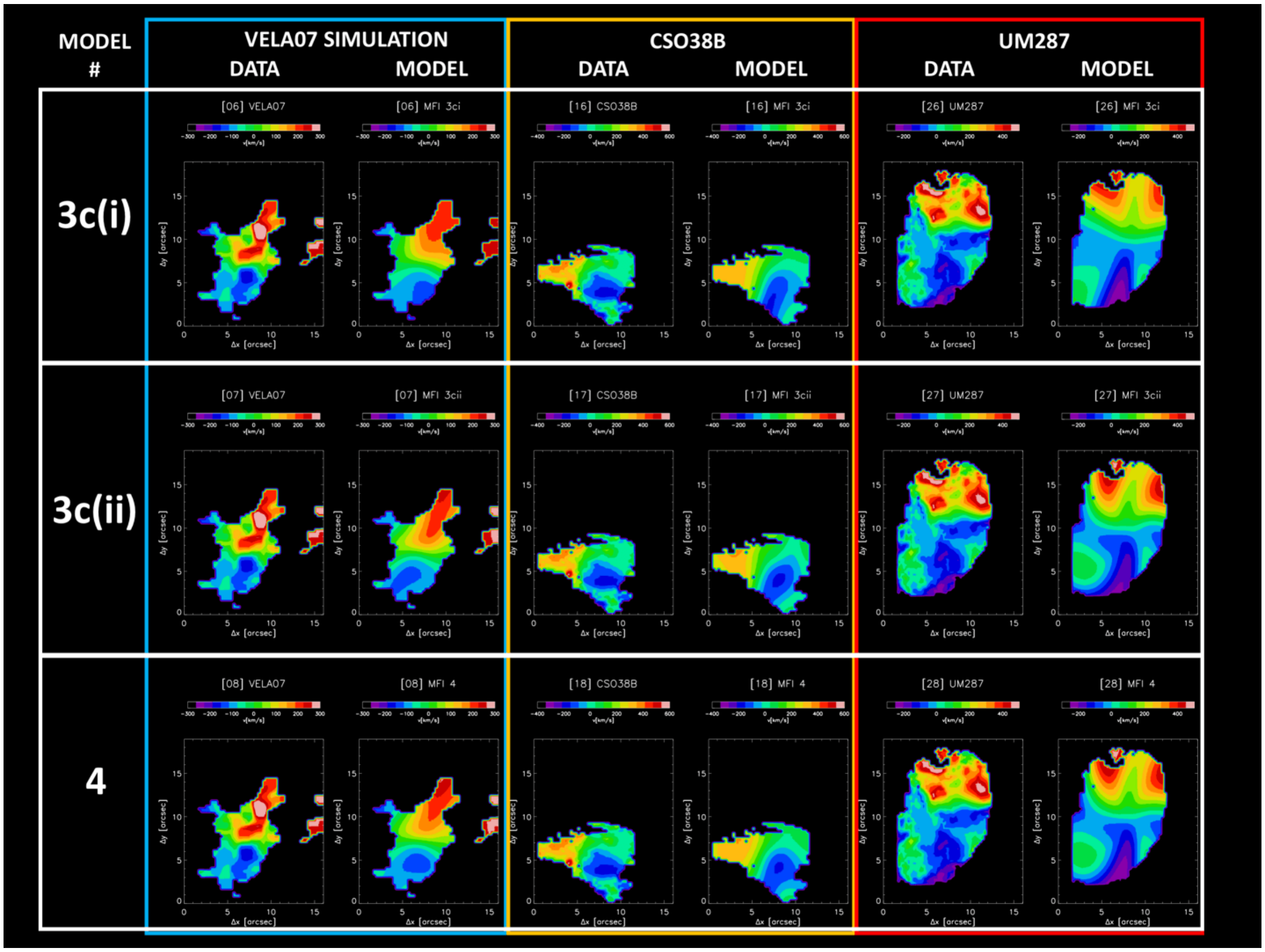

**Supplementary Figure 2c** Velocity maps for the VELA07 simulation (columns 1 and 2), CSO38B (columns 3 and 4), and UM287 (columns 5 and 6). Left-hand image in each column is data, right-hand image is model fit. Models 3c(i), 3c(ii), 4.

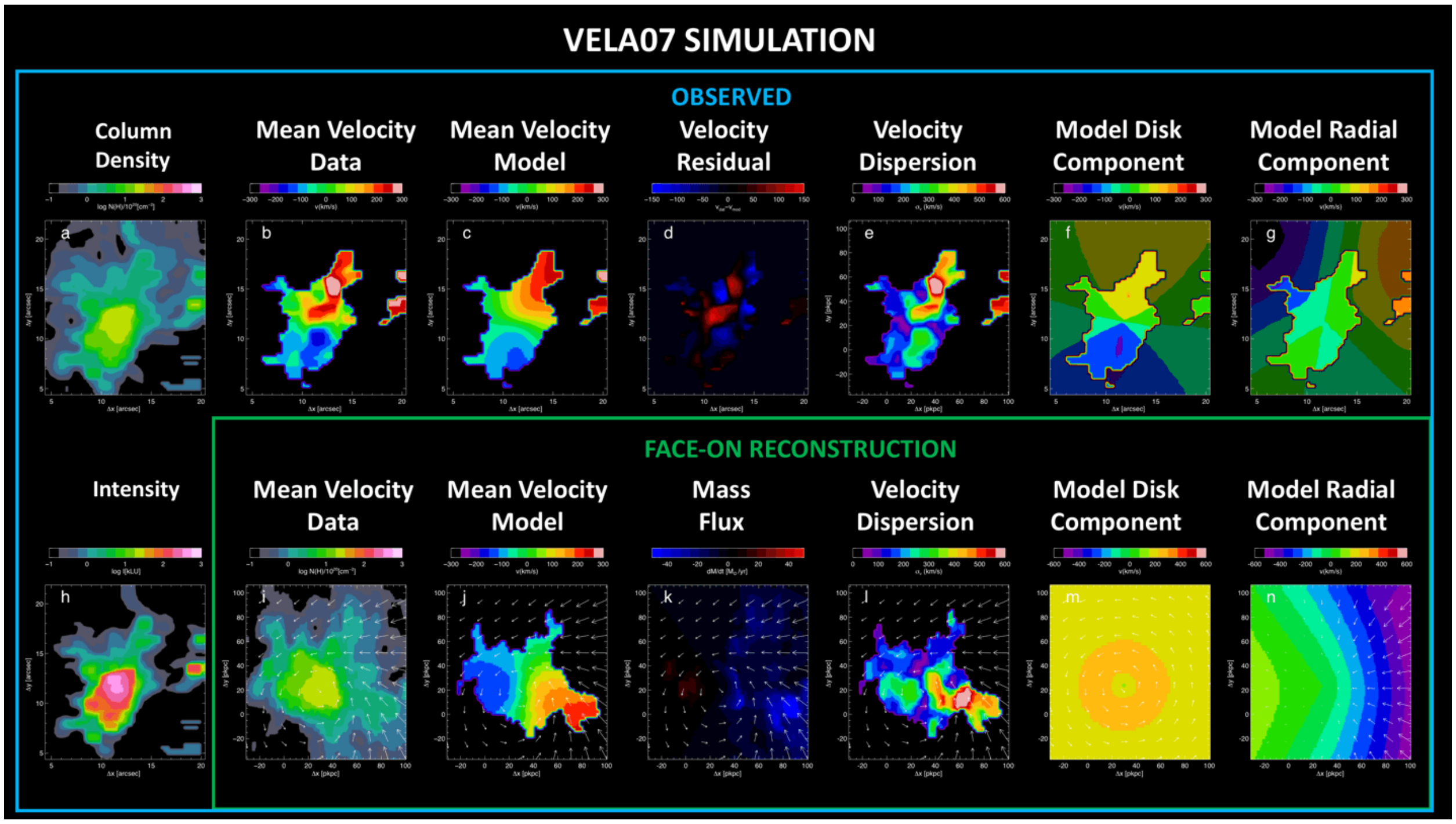

**Supplementary Figure 3** This figure shows a series of panels illustrating the model 3a(i) fit to the VELA07 simulation. a. Inferred column density $N_H$ assuming a thickness of $t_3$=(t/3 kpc)=1 and a clumping factor C=1. The inferred column density can be scaled to other values by multiplying by $(t_3/C)^{1/2}$. b. Mean velocity (data). c. Mean velocity (model). d. $\Delta v = v_{DATA} - v_{MODEL}$. e. velocity dispersion (line width), $\sigma_v$. f. Disk contribution to velocity model, observed frame. g. Radial velocity contribution to velocity model, observed frame. h. Intensity (in kLU = $10^3$ ph cm$^{-2}$ s$^{-1}$ sr$^{-1}$). i. Column density (panel a.) remapped to face-on frame, with velocity field superimposed using vectors. In face-on frame, the line of inclination is in the horizontal direction. The length of the vector in pkpc is the velocity times 0.025. j. Model velocity field remapped to face-on frame without changing velocities, for reference, and face-on velocity vector field. k. radial mass flux in $M_\odot/yr$. The value plotted at each point is $2\pi\, {}^{d\dot{M}}/_{d\phi}$, or the total mass flux through that radius if the value at the point were the same at all azimuth. Face-on velocity vector field superimposed. l. Velocity dispersion remapped to face-on frame, and face-on velocity vector field. There is some trend for velocity dispersion to increase in regions with large velocity gradients, possibly due to multiple components superposed. m. Disk velocity model, face-on frame, and velocity vector field. n. Radial velocity model, face-on frame, and velocity vector field.

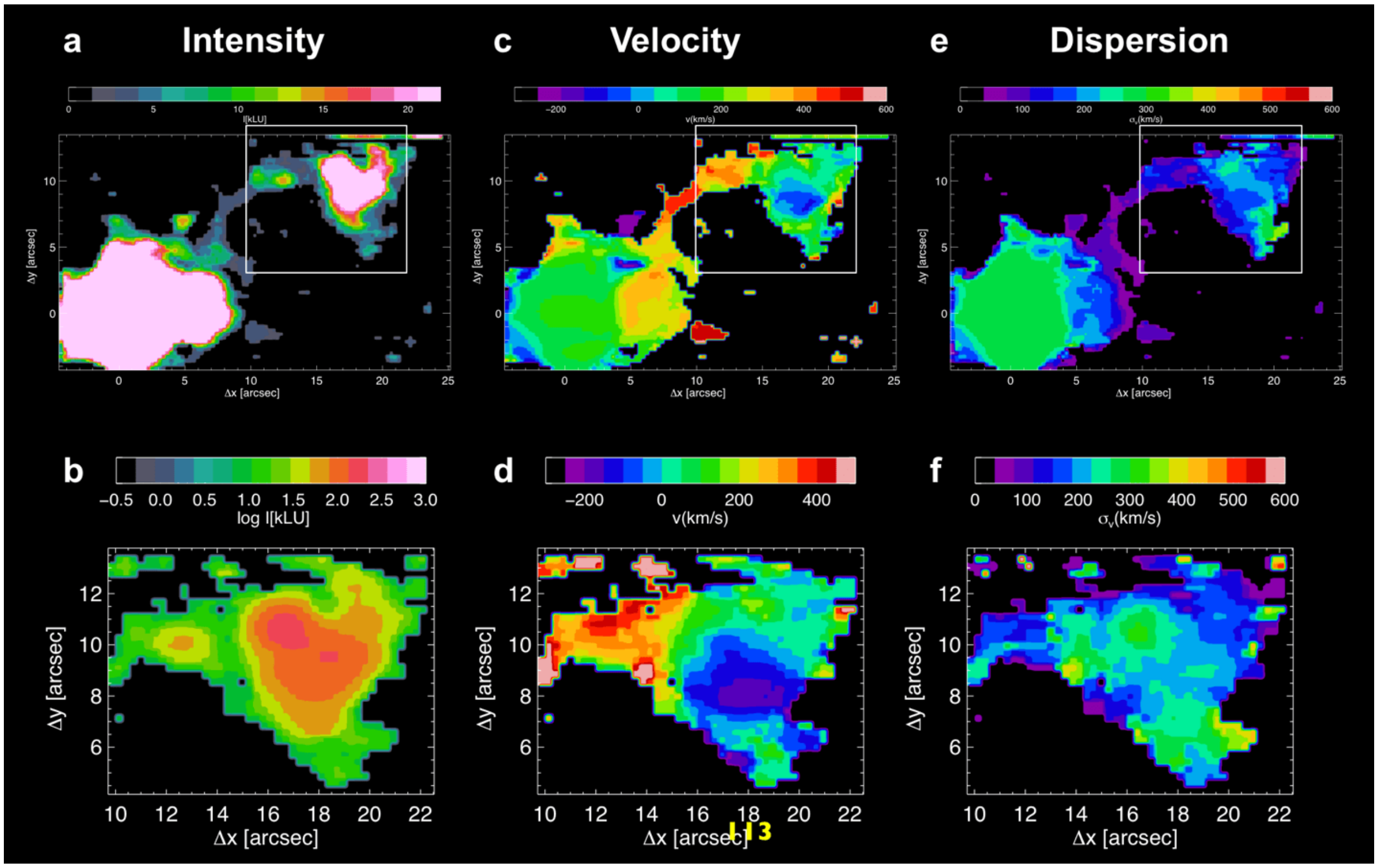

**Supplementary Figure 4** Full field image of CSO38. QSO is in lower left, and Lyα-emitting blob (CSO38B) is in upper right. a. Intensity in narrow band around systemic velocity. b. Zoom in around CSO38B. c. Mean velocity, full field. d. Mean velocity, zoom in around CSO38B. e. Velocity dispersion, full field. f. Velocity dispersion, zoom in around CSO38B.

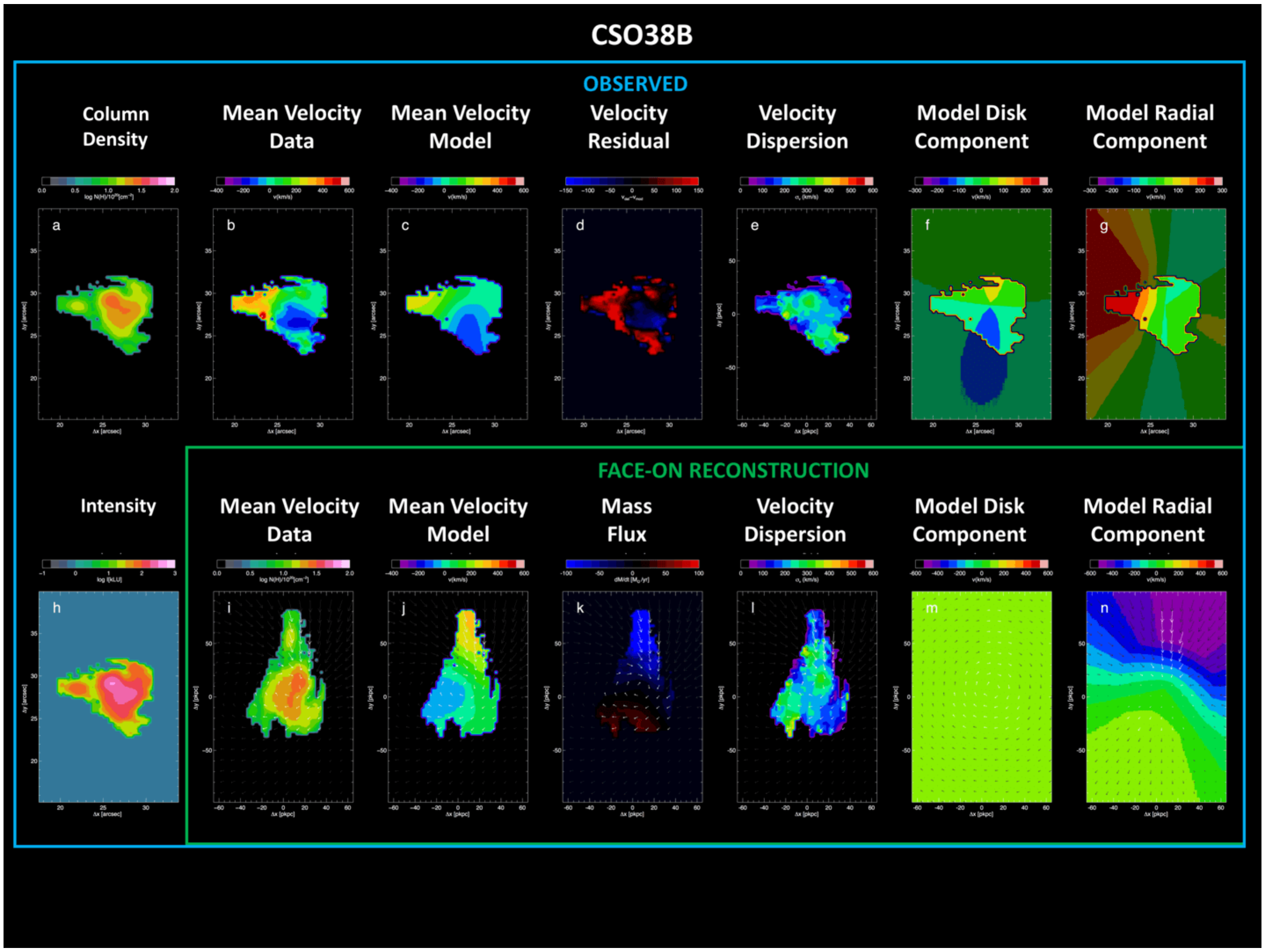

**Supplementary Figure 5** This figure shows a series of panels illustrating the model 3a(i) fit to CSO38B. a. Inferred column density $N_H$ assuming a thickness of $t_3$=(t/3 kpc)=1 and a clumping factor C=1. The inferred column density can be scaled to other values by multiplying by $(t_3/C)^{1/2}$. b. Mean velocity (data). c. Mean velocity (model). d. $\Delta v = v_{DATA} - v_{MODEL}$. e. velocity dispersion (line width), $\sigma_v$. f. Disk contribution to velocity model, observed frame. g. Radial velocity contribution to velocity model, observed frame. h. Intensity (in kLU = $10^3$ ph cm$^{-2}$ s$^{-1}$ sr$^{-1}$). i. Column density (panel a.) remapped to face-on frame, with velocity field superimposed using vectors. In face-on frame, the line of inclination is in the horizontal direction. The length of the vector in pkpc is the velocity times 0.025. j. Model velocity field remapped to face-on frame without changing velocities, for reference, and face-on velocity vector field. k. radial mass flux in $M_\odot/yr$. The value plotted at each point is $2\pi \, {d\dot{M}}/{d\phi}$, or the total mass flux through that radius if the value at the point were the same at all azimuth. Face-on velocity vector field superimposed. l. Velocity dispersion remapped to face-on frame, and face-on velocity vector field. There is some trend for velocity dispersion to increase in regions with large velocity gradients, possibly due to multiple components superposed. m. Disk velocity model, face-on frame, and velocity vector field. n. Radial velocity model, face-on frame, and velocity vector field.

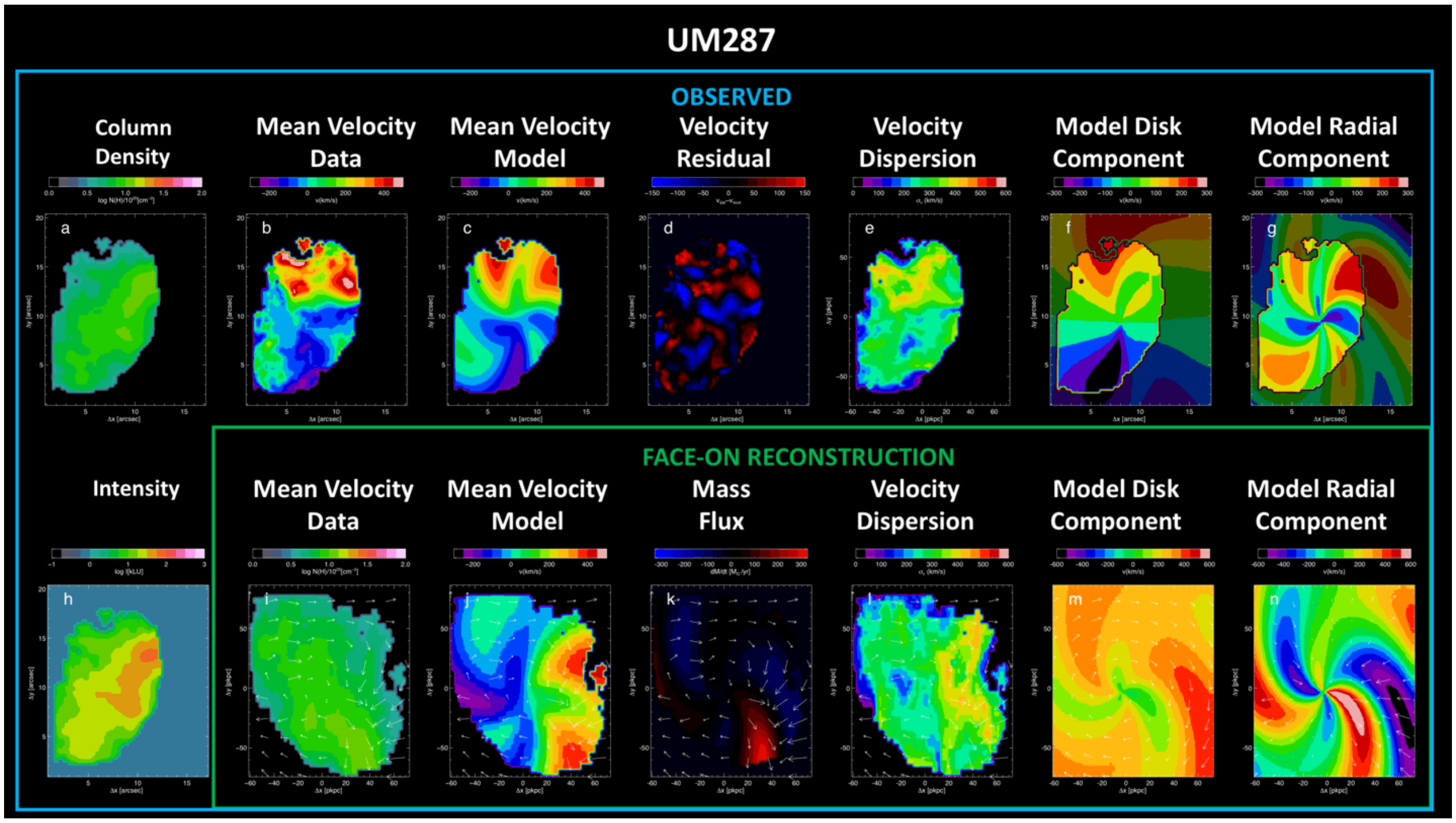

**Supplementary Figure 6** This figure shows a series of panels illustrating the model 4 fit to UM287. a. Inferred column density $N_H$ assuming a thickness of $t_3$=(t/3 kpc)=1 and a clumping factor C=1. The inferred column density can be scaled to other values by multiplying by $(t_3/C)^{1/2}$. b. Mean velocity (data). c. Mean velocity (model). d. $\Delta v = v_{DATA} - v_{MODEL}$. e. velocity dispersion (line width), $\sigma_V$. f. Disk contribution to velocity model, observed frame. g. Radial velocity contribution to velocity model, observed frame. h. Intensity (in kLU = $10^3$ ph cm$^{-2}$ s$^{-1}$ sr$^{-1}$). i. Column density (panel a.) remapped to face-on frame, with velocity field superimposed using vectors. In face-on frame, the line of inclination is in the horizontal direction. The length of the vector in pkpc is the velocity times 0.025. j. Model velocity field remapped to face-on frame without changing velocities, for reference, and face-on velocity vector field. k. radial mass flux in $M_\odot/yr$. The value plotted at each point is $2\pi \, {}^{d\dot{M}}/_{d\phi}$, or the total mass flux through that radius if the value at the point were the same at all azimuth. Face-on velocity vector field superimposed. l. Velocity dispersion remapped to face-on frame, and face-on velocity vector field. There is some trend for velocity dispersion to increase in regions with large velocity gradients, possibly due to multiple components superposed. m. Disk velocity model, face-on frame, and velocity vector field. n. Radial velocity model, face-on frame, and velocity vector field.

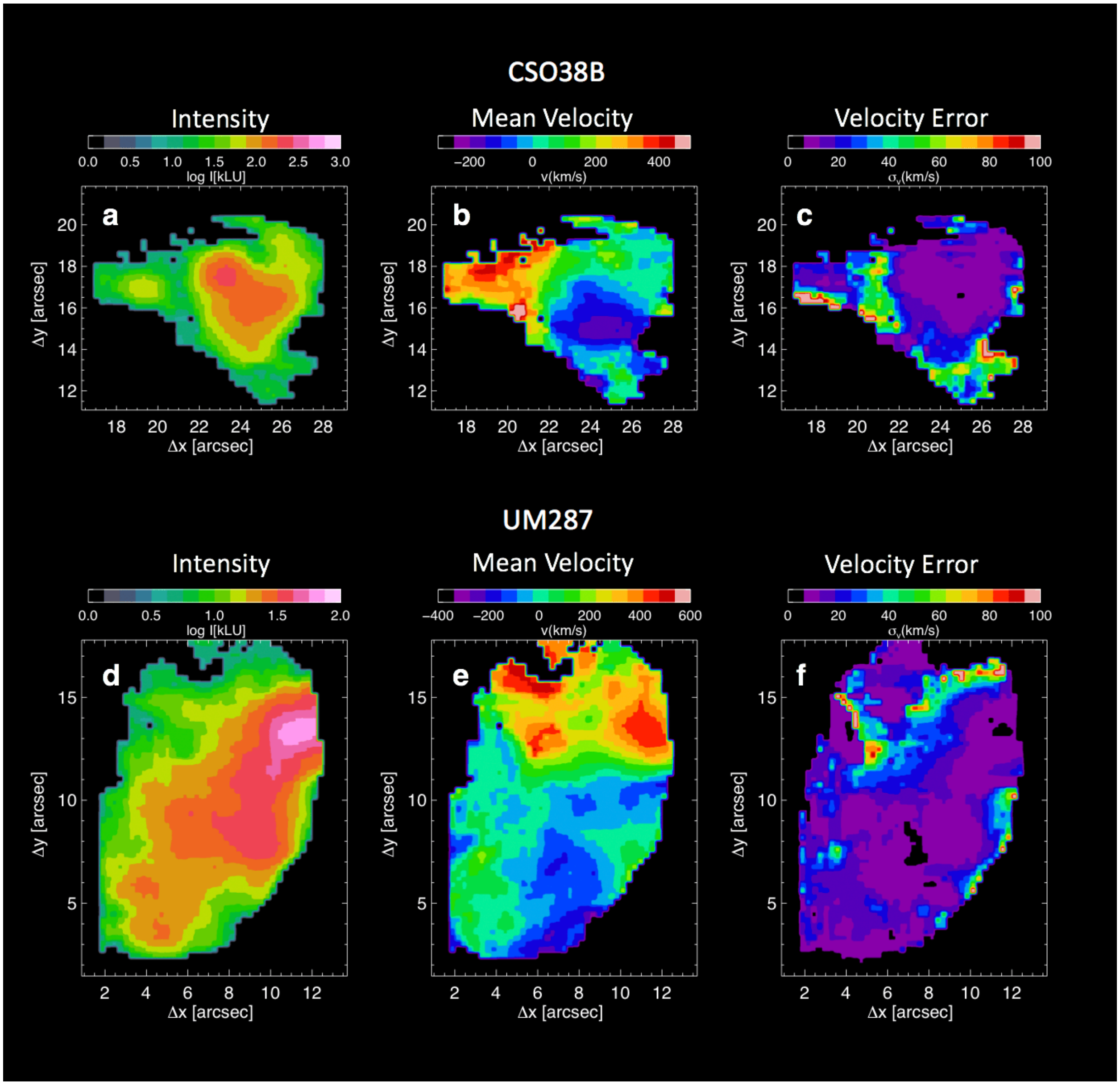

**Supplementary Figure 7** – Narrow-band image, mean velocity map and velocity centroid error for CSO38B (a-c) and UM287 (d-f). Note scale of velocity error is expanded by a factor of 10 over that of the mean velocity.

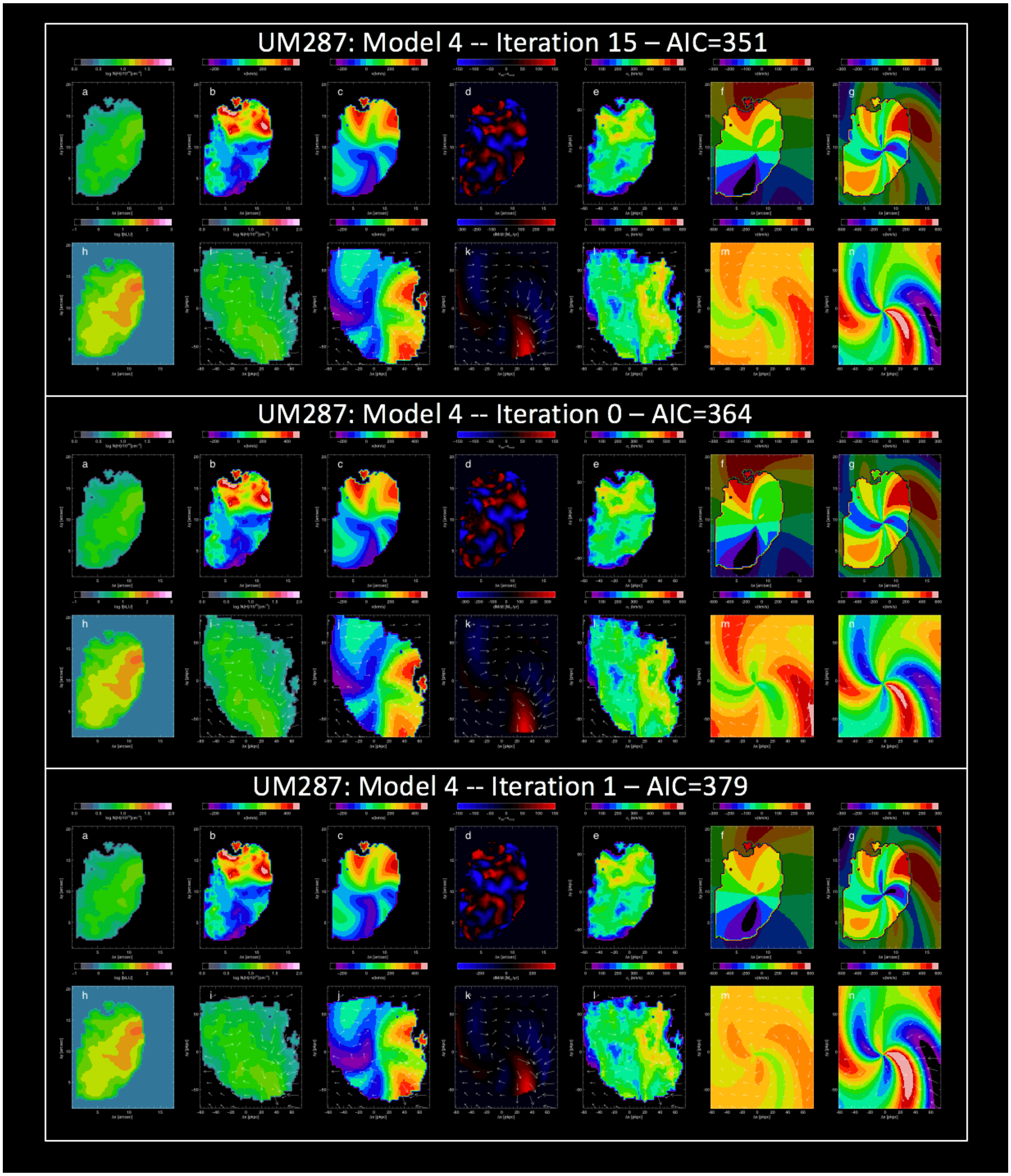

**Supplementary Figure 8** – Three iterations with minimum $\chi^2$ for object UM287, Model 4. Individual panels for each iteration are the same as in Supplementary Fig. 3, 5, and 6. Iteration 15 is identical to Supplementary Fig. 6. The MFI components (panel n in each iteration) are almost identical for these three local minima. In other words, the MFI component is not suppressed in nearby local minima of the optimization function.

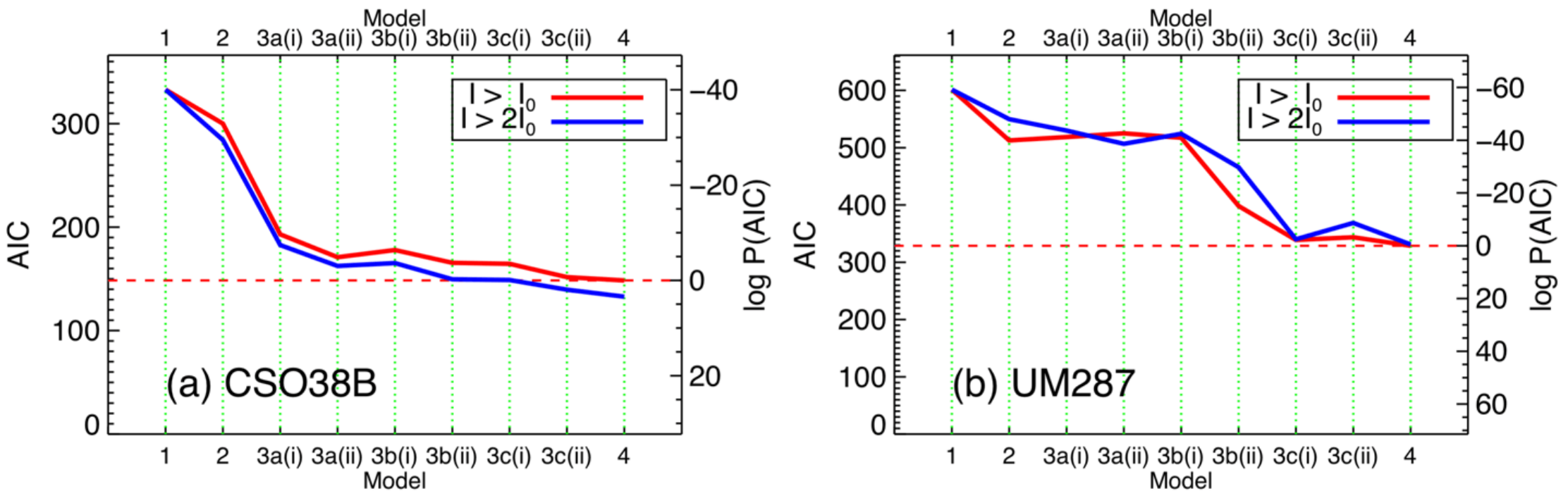

**Supplementary Figure 9** – AIC vs. Model for two SNR thresholds. a. CSO38B, red line shows baseline SNR threshold ($I_0$ = 7 kLU), blue line shows SNR threshold twice as large, ($2I_0$ = 14 kLU). The dramatic drop in AIC for Model 3a(i) is unchanged. b. UM287, red line shows baseline SNR threshold ($I_0$ = 4 kLU), blue line shows SNR threshold twice as large, ($2I_0$ = 8 kLU). The pronounced drop in AIC for model 3c(i) remains.

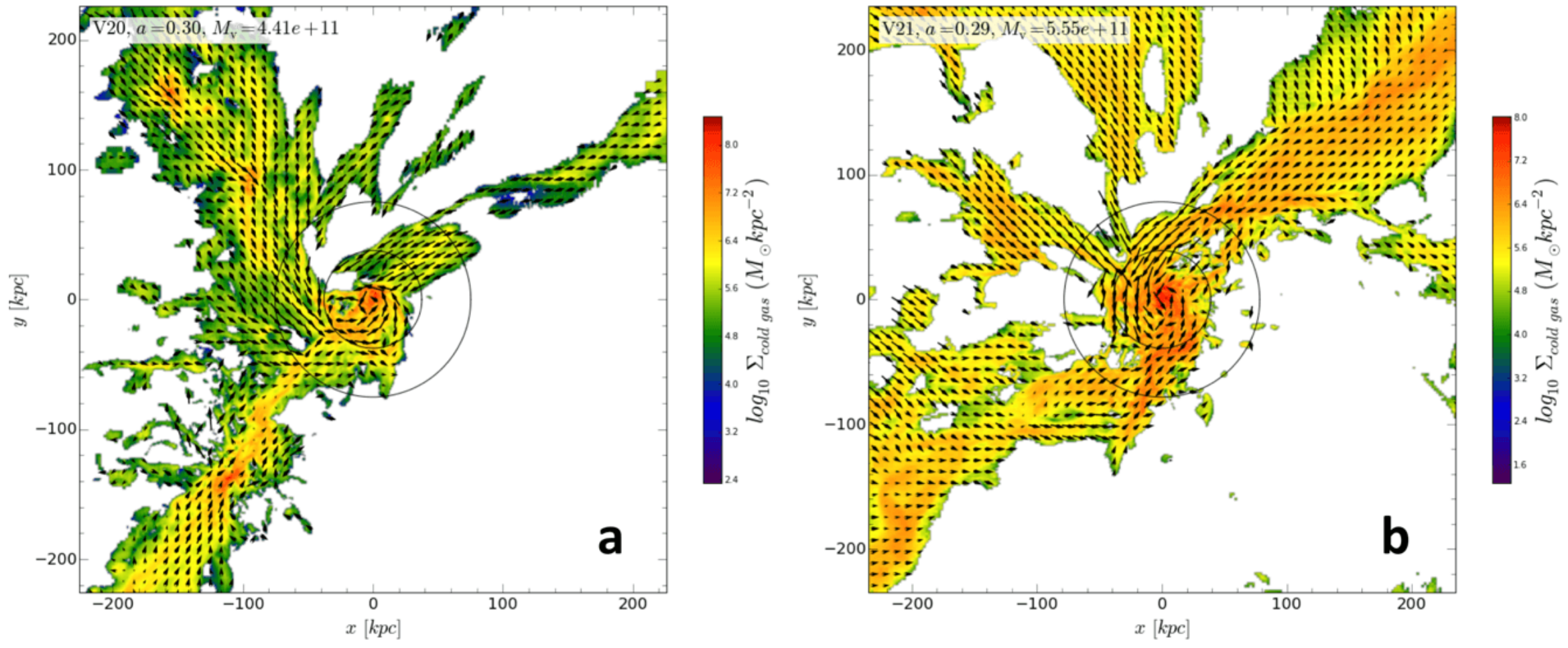

**Supplementary Figure 10** – Column density and velocity flow field for simulated galaxies VELA20 and VELA21. Compare to VELA07, Fig. 1.

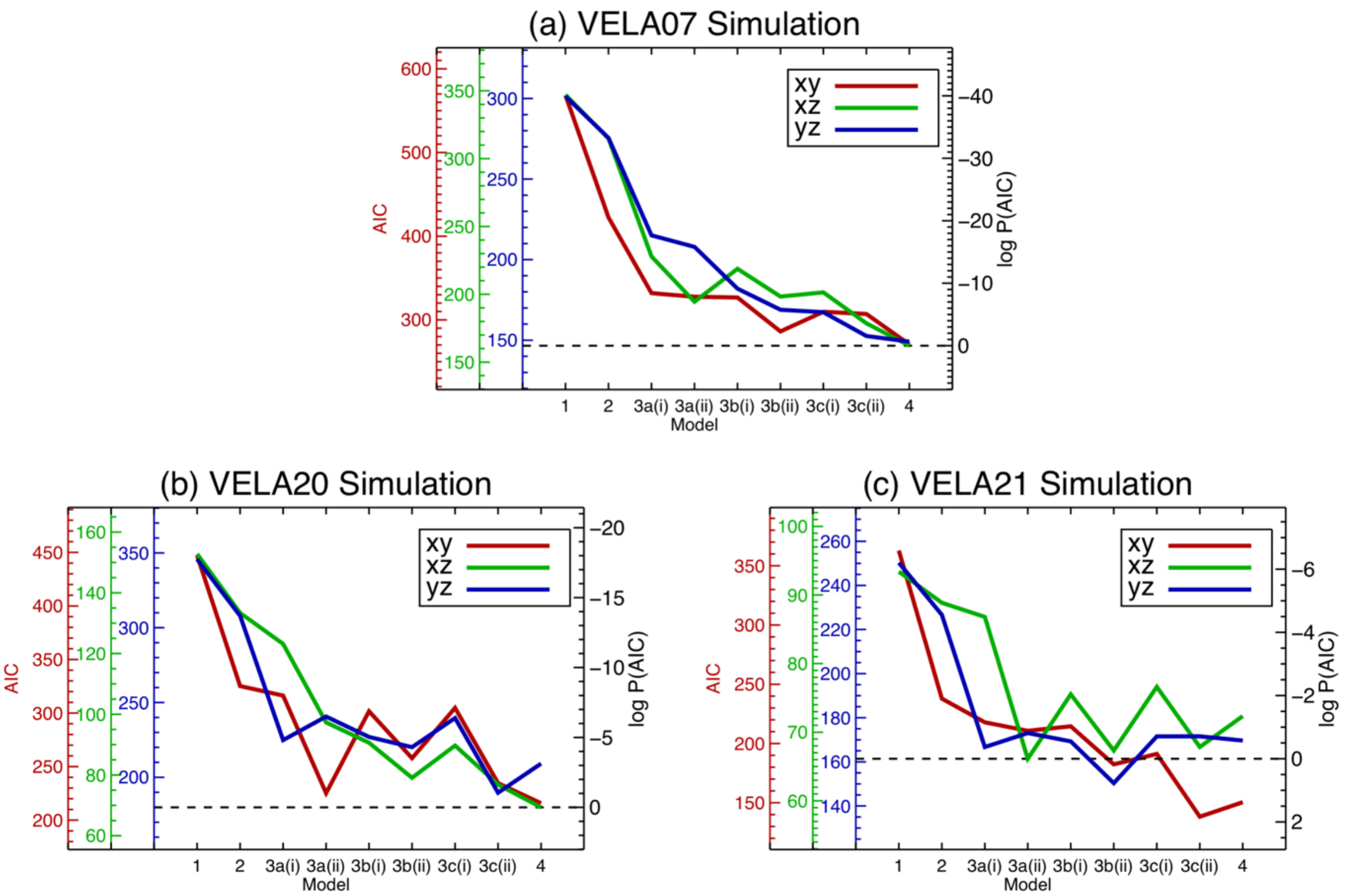

**Supplementary Figure 11** – AIC vs. Model for three line of sight views of simulated galaxies: (a) VELA07 simulation, (b) VELA20 simulation, and (c) VELA21 simulation. Left hand axis shows the AIC for each view, coded by color. Right hand axis shows the log of the formal probability that the higher AIC model is a true one, in each case for the model with the smallest change in AIC, which in all three cases is the xz view. Horizontal dashed line corresponds to the minimum AIC for the xz model.

## Supplementary Tables

### Supplementary Table 1 – Summary of Observations

| **Target Name(s)** | UM287 / QSO J0052+0101 | CSO38 / QSO B1009+2956 |
|---|---|---|
| **Coordinates** | 00:52:02.400 +01:01:29.300 | 10:11:56.000 +29:41:42.000 |
| **Redshift** | 2.28 | 2.652 |
| **Image Slicer** | Medium (M) | Large (L) |
| **Grating** | Blue Medium Resolution (BM) | Blue Medium Resolution (BM) |
| **Central Wavelength** | 4290Å | 4500Å |
| **Resolution** | 5000 | 2500 |
| **Source Exposure Time** | 4h (12×20m) | 2h (12×5m, 3×20m) |
| **Sky Exposure Time** | 50m (5×10m) | 10m (1x10m) |
| **Sky Subtraction Method** | Interleaved sky exposures | In-field sky measurement |
| **Date(s) of Observation** | 17th October 2017 | 15th April 2017, 22nd November 2018 |
| **Sensitivity Achieved (5σ)** | $3 \times 10^{-18}$ erg cm$^{-2}$ s$^{-1}$ arcsec$^{-2}$ (1″x1″)<br>30 kLU | $2.8 \times 10^{-18}$ erg cm$^{-2}$ s$^{-1}$ arcsec$^{-2}$ (1″x1″)<br>28 kLU |

### Supplementary Table 2 – Revised MFI Criteria and Comparison to CSO38B and UM287 nebulae.

| **Criterion** | **CSO38B Nebula** | **UM287 Nebula** |
|---|---|---|
| 1. Higher intensity located *approximately symmetrically* around 1D velocity center and relatively uniform intensity with clear intensity break at edges | Yes. Highest intensity located at maximum velocity shear. | Yes. Highest intensity located at maximum velocity shear. |
| 3. 2D velocity and intensity distribution consistent with a disk *and multi-filament radial flow* in an NFW halo with minimal residuals | Yes based on fits. | Yes based on fits. |
| 4. Evidence for one or more filaments with low velocity gradient and possibly lower-velocity dispersion *aligned with inferred radial flow direction(s)*; | Yes. | Yes. |
| 5. *Velocity gradients transverse to rotation-induced velocity shear.* | Yes. | Yes. |
| 6. Star formation *near* center of disk *with radial mass flux onto galaxy consistent with star formation rate.* | Yes, object BX173. | Yes, object C, D, and E. |
| *7. Object is separated from illuminating QSO and does not appear to be part of interaction producing the QSO.* | Yes. | Yes. |
| **Summary** | **6/6 criteria matched** | **6/6 criteria matched.** |

**Supplementary Table 3 – Kinematic Models with Multi-Filament Inflow**

| Model | Cartoon | Formula |
|---|---|---|
| **1**<br>**Rotation** | Keplerian disk rotation in an NFW dark matter halo | |
| | 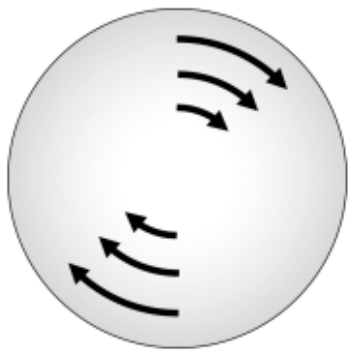 | $v_\phi(r) = \sqrt{\frac{GM(r; M_h, c)}{r}}$ |
| | | $v_r(r) = 0$ |
| **2**<br>**Linear radial** | Disk plus radially linear radial flow | |
| | 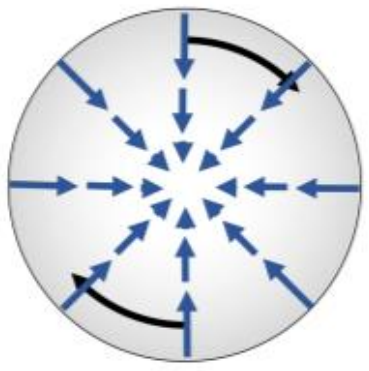 | $v_\phi(r) = \sqrt{\frac{GM(r; M_h, c)}{r}}$ |
| | | $v_r(r, \phi) = v_{r0}\left(\frac{r}{r_V}\right)$ |
| **3a**<br>**MFI 1 mode**<br>**i no spiral**<br>**ii spiral** | Disk plus azimuthally modulated linear radial flow, mode 1. | |
| | 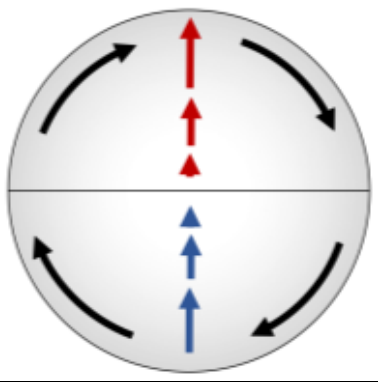 | $v_\phi(r) = \sqrt{\frac{GM(r; M_h, c)}{r}}$ |
| | | $v_r(r, \phi) = \left(\frac{r}{r_V}\right)\{v_{r0} + v_{r1}\sin(\phi + \phi_1 + a_{sp} r)\}$ |
| **3b**<br>**MFI 2 mode**<br>**i no spiral**<br>**ii spiral** | Disk plus azimuthally modulated linear radial flow, mode 1+2. | |
| | 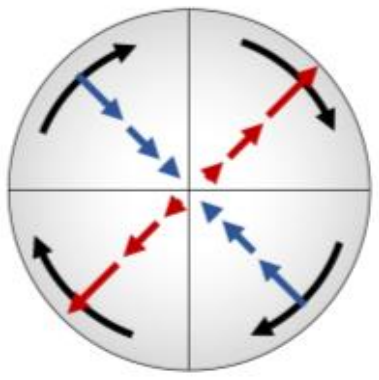 | $v_\phi(r) = \sqrt{\frac{GM(r; M_h, c)}{r}}$ |
| | | $v_r(r, \phi) = \left(\frac{r}{r_V}\right)\left\{v_{r0} + \sum_{n=1}^{2} v_{rn}\sin(n\phi + \phi_n + a_{sp} r)\right\}$ |
| **3c**<br>**MFI 3 mode**<br>**i no spiral**<br>**ii spiral** | Disk plus azimuthally modulated linear radial flow, mode 1+2+3. | |
| | 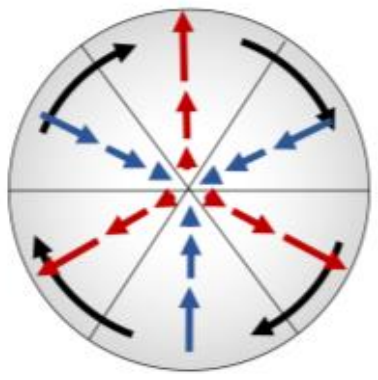 | $v_\phi(r) = \sqrt{\frac{GM(r; M_h, c)}{r}}$ |
| | | $v_r(r, \phi) = \left(\frac{r}{r_V}\right)\left\{v_{r0} + \sum_{n=1}^{3} v_{rn}\sin(n\phi + \phi_n + a_{sp} r)\right\}$ |
| **4**<br>**MFI 3 mode**<br>**radial +**<br>**azimuthal** | Disk plus azimuthally modulated linear radial + azimuthal flow, mode 1+2+3. | |
| | 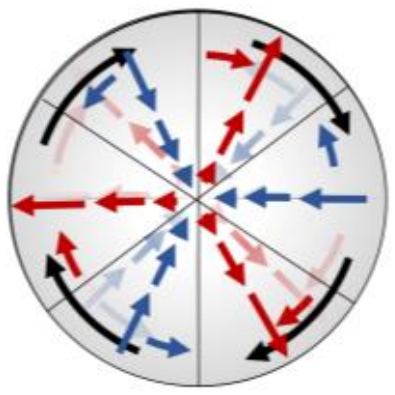 | $v_\phi(r) = \sqrt{\frac{GM(r; M_h, c)}{r}} + f_a v_r(r, \phi)$ |
| | | $v_r(r, \phi) = \left(\frac{r}{r_V}\right)\left\{v_{r0} + \sum_{n=1}^{3} v_{rn}\sin(n\phi + \phi_n + a_{sp} r)\right\}$ |

**Supplementary Table 4 – VELA07 Simulation MFI Fit Parameters**

| **Param** | **Model** | | | | | | | | |
|---|---|---|---|---|---|---|---|---|---|
| | **1** | **2** | **3a(i)** | **3a(ii)** | **3b(i)** | **3b(ii)** | **3c(i)** | **3c(ii)** | **4** |
| **#params** | 5 | 6 | 8 | 9 | 10 | 11 | 12 | 13 | 14 |
| $\chi^2$ | 372 | 272 | 208 | 204 | 204 | 176 | 188 | 156 | 152 |
| **DOF** | 163 | 162 | 160 | 159 | 157 | 156 | 154 | 153 | 152 |
| **AIC** | 196 | 148 | 121 | 121 | 125 | 114 | 125 | 111 | 112 |
| **P(AIC)** | 4e-19 | 1e-8 | 0.01 | 0.01 | 0.001 | 0.31 | 0.001 | 1 | 1 |
| **BIC** | 397 | 302 | 249 | 250 | 255 | 232 | 249 | 222 | 223 |
| **P(BIC)** | 1.6e-38 | 4e-18 | 2.2e-6 | 8.3e-7 | 1.1e-7 | 7e-3 | 2e-6 | 1 | 1 |
| $\sigma_V$ | 74 | 64 | 56 | 55 | 55 | 51 | 53 | 48 | 47 |
| **Log $M_h$** | 12.81 | 11.83 | 11.89 | 11.81 | 12.00 | 11.88 | 11.96 | 11.91 | 11.98 |
| **Log c** | 0.02 | 1.00 | 1.00 | 1.00 | 1.00 | 1.00 | 1.00 | 1.00 | 1.00 |
| **Inc** | 40 | 40 | 40 | 46 | 40.1 | 55 | 40.1 | 54.7 | 53.1 |
| $\phi_0$ | 50 | 84 | 80 | 76 | 85 | 67 | 85 | 63 | 76.9 |
| $v_{r0}$ | 0 | -276 | -198 | -120 | -244 | -48 | -196 | 4.4 | -96 |
| $a_{sp}$ | -- | -- | -- | 0.62 | -- | -3.9 | -- | -3.6 | -3.9 |
| $\phi_1$ | -- | -- | 3.6 | 39 | -5 | -70 | 20 | 9 | 0 |
| $v_{r1}$ | -- | -- | -380 | -314 | -336 | -144 | -504 | -232 | -210 |
| $\phi_2$ | -- | -- | -- | -- | -306 | -30 | 54 | -18 | -30 |
| $v_{r2}$ | -- | -- | -- | -- | -157 | -177 | -63 | -201 | -158 |
| $\phi_3$ | -- | -- | -- | -- | -- | -- | 31 | 19 | 9 |
| $v_{r3}$ | -- | -- | -- | -- | -- | -- | -224 | -194 | -173 |
| $f_a$ | -- | -- | -- | -- | -- | -- | -- | -- | 0.54 |
| **$dM/dt_{max}$** | 0 | -356 | -164 | -123 | -188 | -188 | -274 | -184 | -250 |
| $r_{dM/dt(max)}$ | -- | 30 | 60 | 70 | 60 | 60 | 60 | 45 | 35 |
| **<dM/dt>(r<50kpc)** | 0 | -37 | -24 | -14 | -15 | -6 | -21 | -7 | -24 |
| $R_V$ | 177 | 83 | 87 | 82 | 95 | 87 | 92 | 89 | 94 |
| $v_c$ | 400 | 187 | 196 | 184 | 213 | 195 | 207 | 199 | 210 |
| $M_b$ | 2.9e11 | 3.0e11 | 3.0e11 | 3.3e11 | 3.0e11 | 4.0e11 | 3.0e11 | 3.9e11 | 3.8e11 |
| $M_b/M_d$ | 0.046 | 0.44 | 0.39 | 0.51 | 0.30 | 0.52 | 0.33 | 0.48 | 0.40 |
| $\langle V_a \rangle$ | 282 | 103 | 108 | 102 | 118 | 107 | 114 | 110 | 110 |
| $\langle V_r \rangle$ | 0 | -51 | -24 | -20 | -26 | -26 | -43 | -18 | -27 |
| $L_b$ | 1.3e15 | 1.0e14 | 1.0e15 | 1.1e15 | 1.1e15 | 1.6e15 | 1.1e15 | 1.6e15 | 1.3e15 |
| $j_b$ | 4.2e3 | 3.3e3 | 3.5e3 | 3.4e3 | 3.8e3 | 4.0e3 | 3.7e3 | 4.1e3 | 3.5e3 |
| $\lambda_b$ | 0.06 | 0.21 | 0.20 | 0.22 | 0.19 | 0.23 | 0.19 | 0.23 | 0.18 |
| | | | | | | | | | |

**Supplementary Table 5 – CSO38B Fit Parameters**

| **Param** | **Model** | | | | | | | | |
|---|---|---|---|---|---|---|---|---|---|
| | **1** | **2** | **3a(i)** | **3a(ii)** | **3b(i)** | **3b(ii)** | **3c(i)** | **3c(ii)** | **4** |
| **#params** | 5 | 6 | 8 | 9 | 10 | 11 | 12 | 13 | 14 |
| $\chi^2$ | 331 | 288 | 177 | 171 | 163 | 155 | 152 | 142 | 138 |
| **DOF** | 140 | 139 | 137 | 136 | 134 | 133 | 131 | 130 | 129 |
| **AIC** | 341 | 300 | 194 | 190 | 187 | 181 | 183 | 176 | 174 |
| **P(AIC)** | 5.2e-37 | 3.9e-28 | 5.4e-5 | 3.5e-4 | 0.002 | 0.03 | 0.01 | 0.49 | 1.0 |
| **BIC** | 356 | 318 | 217 | 215 | 222 | 209 | 211 | 206 | 207 |
| **P(BIC)** | 3e-33 | 6e-25 | 0.005 | 0.009 | 0.04 | 0.19 | 0.07 | 1 | 0.55 |
| $\sigma_v$ | 110 | 104 | 82 | 80 | 78 | 76 | 76 | 73 | 72 |
| **Log** $M_h$ | 13.0 | 11.96 | 11.19 | 11.12 | 11.52 | 11.37 | 11.50 | 11.71 | 11.66 |
| **Log c** | 0.33 | 1.00 | 1.00 | 1.00 | 1.00 | 1.00 | 1.00 | 1.00 | 1.00 |
| **Inc** | 34 | 34 | 40 | 30 | 40.1 | 67 | 68 | 66 | 70 |
| $\phi_0$ | 145 | 103 | 80 | 63 | 85 | 68 | 78 | 76 | 98 |
| $v_{r0}$ | -- | -517 | -173 | -486 | -244 | -71 | -115 | -314 | 43 |
| $a_{sp}$ | -- | -- | -- | 3.0 | -- | 1.0 | -- | -2.8 | -0.53 |
| $\phi_1$ | -- | -- | 89 | 80 | -5 | 12 | -118 | -89 | -162 |
| $v_{r1}$ | -- | -- | -318 | 608 | -336 | 270 | 530 | -565 | 350 |
| $\phi_2$ | -- | -- | -- | -- | -306 | 54 | 15 | 282 | -23 |
| $v_{r2}$ | -- | -- | -- | -- | -157 | 147 | -68 | -166 | 186 |
| $\phi_3$ | -- | -- | -- | -- | -- | -- | 17 | -209 | -18 |
| $v_{r3}$ | -- | -- | -- | -- | -- | -- | -205 | 330 | -62 |
| $f_a$ | -- | -- | -- | -- | -- | -- | -- | -- | 1.12 |
| $dM/dt_{max}$ | 0 | -243 | -117 | -346 | -29 | -82 | -74 | -227 | -80 |
| $r_{dM/dt(max)}$ | -- | 30 | 30 | 25 | 35 | 45 | 35 | 30 | 110 |
| **<dM/dt>** | 0 | -37 | -39 | -64 | -30 | -23 | -35 | -37 | -8 |
| $R_V$ | 204 | 92 | 51 | 82 | 66 | 87 | 65 | 76 | 73 |
| $v_c$ | 460 | 207 | 114 | 184 | 148 | 195 | 145 | 170 | 164 |
| $M_b$ | 5.2e10 | 5.1e10 | 7.7e10 | 5.0e10 | 1.1e11 | 1.1e11 | 1.2e11 | 1.1e11 | 1.3e11 |
| $M_b/M_d$ | 0.005 | 0.06 | 0.50 | 0.38 | 0.36 | 0.48 | 0.37 | 0.21 | 0.27 |
| $\langle V_a \rangle$ | 113 | 236 | 122 | 120 | 122 | 137 | 143 | 185 | 197 |
| $\langle V_r \rangle$ | 0 | -136 | -80 | -240 | -46 | -68 | -49 | -77 | -20 |
| $L_b$ | 3.9e14 | 3.1e14 | 2.9e14 | 1.4e14 | 8.1e14 | 6.4e14 | 7.7e14 | 8.1e14 | 1.5e15 |
| $j_b$ | 7.5e3 | 4.9e3 | 3.8e3 | 2.8e3 | 3.6e3 | 5.8e3 | 6.6e3 | 7.6e3 | 1.2e4 |
| $\lambda_b$ | 0.08 | 0.31 | 0.66 | 0.54 | 0.68 | 0.75 | 0.71 | 0.59 | 0.95 |

**Supplementary Table 6 – UM287 Fit Parameters**

| Param | Model | | | | | | | | |
|---|---|---|---|---|---|---|---|---|---|
| | **1** | **2** | **3a(i)** | **3a(ii)** | **3b(i)** | **3b(ii)** | **3c(i)** | **3c(ii)** | **4** |
| **#params** | 5 | 6 | 8 | 9 | 10 | 11 | 12 | 13 | 14 |
| $\chi^2$ | 755 | 723 | 577 | 512 | 527 | 483 | 403 | 321 | 317 |
| **DOF** | 312 | 311 | 309 | 308 | 306 | 305 | 303 | 302 | 301 |
| **AIC** | 765 | 735 | 593 | 530 | 549 | 508 | 432 | 352 | 351 |
| **P(AIC)** | 0 | 0 | <1e-30 | <1e-30 | <1e-30 | <1e-30 | <1e-10 | 0.4 | 1 |
| **BIC** | 784 | 757 | 622 | 563 | 584 | 546 | 471 | 395 | 397 |
| **P(BIC)** | 1e-84 | 5e-79 | 9e-50 | 6e-37 | 2e-41 | 4e-33 | 5e-17 | 1 | 0.8 |
| $\sigma_v$ | 111 | 109 | 97 | 92 | 93 | 89 | 81 | 72 | 72 |
| **Log $M_h$** | 12.71 | 12.20 | 11.30 | 11.45 | 11.06 | 11.45 | 12.52 | 12.50 | 12.69 |
| **Log c** | 0.0 | 0.33 | 0.0 | 0.0 | 1.0 | 0.0 | 0.0 | 0.16 | 0.0 |
| **Inc** | 65 | 70 | 50 | 57 | 61 | 62 | 59 | 51 | 54 |
| $\phi_0$ | 92 | 113 | 159 | 147 | 184 | 135 | 85 | 82 | 84 |
| $v_{r0}$ | 0 | -114 | -245 | -205 | -16 | -126 | -181 | -70 | -108 |
| $a_{sp}$ | 0 | 0 | 0 | 1.7 | 0 | 2.7 | 0 | 4.7 | 4.1 |
| $\phi_1$ | 0 | 0 | -86 | 45 | 72 | -22 | 47 | -165 | -20 |
| $v_{r1}$ | 0 | 0 | -236 | -218 | 220 | 161 | 802 | 773 | -956 |
| $\phi_2$ | 0 | 0 | 0 | 0 | -197 | 117 | -178 | -52 | -55 |
| $v_{r2}$ | 0 | 0 | 0 | 0 | 257 | 114 | -187 | 208 | 214 |
| $\phi_3$ | 0 | 0 | 0 | 0 | 0 | 0 | -38 | 6 | 54 |
| $v_{r3}$ | 0 | 0 | 0 | 0 | 0 | 0 | -718 | -920 | 1050 |
| $f_a$ | 0 | 0 | 0 | 0 | 0 | 0 | 0 | 0 | 0.06 |
| $dM/dt_{max}$ | 0 | 0 | 0 | 0 | 0 | 0 | -180 | -240 | -300 |
| $r_{dM/dt(max)}$ | 0 | 0 | 0 | 0 | 0 | 0 | 60 | 60 | 70 |
| **<dM/dt>** | 0 | -54 | -79 | -78 | -69 | -83 | -17 | -20 | -26 |
| $R_V$ | 165 | 112 | 56 | 63 | 46 | 63 | 143 | 140 | 162 |
| $v_c$ | 367 | 248 | 124 | 139 | 103 | 139 | 317 | 312 | 361 |
| $M_b$ | 1.5E+11 | 1.8E+11 | 9.8E+10 | 1.2E+11 | 1.3E+11 | 1.3E+11 | 1.2E+11 | 1.0E+11 | 1.1E+11 |
| $M_b/M_d$ | 0.03 | 0.11 | 0.49 | 0.41 | 1.13 | 0.48 | 0.04 | 0.03 | 0.02 |
| $\langle V_a \rangle$ | 255 | 219 | 109 | 122 | 95 | 124 | 220 | 227 | 235 |
| $\langle V_r \rangle$ | 0 | -66 | -186 | -160 | -147 | -125 | 5 | -10 | -6 |
| $L_b$ | 2.3E+15 | 2.9E+15 | 5.5E+14 | 8.3E+14 | 6.8E+14 | 1.1E+15 | 1.4E+15 | 1.0E+15 | 1.2E+15 |
| $j_b$ | 1.6E+04 | 1.6E+04 | 5.6E+03 | 7.2E+03 | 5.2E+03 | 8.1E+03 | 1.1E+04 | 1.0E+04 | 1.1E+04 |
| $\lambda_b$ | 0.26 | 0.59 | 0.81 | 0.82 | 1.09 | 0.93 | 0.25 | 0.24 | 0.19 |

**Supplementary Table 7 – Best Fit Model Parameter Error Limits**

| Object | VELA07 Simulation | CSO38B | UM287 |
|---|---|---|---|
| **Model** | **3a(i)** | **3a(i)** | **4** |
| **Log $M_h$** | $11.90^{+0.19}_{-0.09}$ | $11.25^{+0.11}_{-0.16}$ | $12.69^{+0.04}_{-0.16}$ |
| **Log c** | 1.00 | 1.00 | 0.0 |
| **Inc** | $40^{+6}_{-6}$ | $40^{+4.8}_{-5.0}$ | $54^{+2}_{-2}$ |
| **$\phi_0$** | $80^{+5}_{-5}$ | $80^{+4}_{-6}$ | $84^{+1.4}_{-2.9}$ |
| **$v_{r0}$** | $-209^{+65}_{-76}$ | $-173^{+23}_{-21}$ | $-108^{+46}_{-19}$ |
| **$a_{sp}$** | -- | -- | $4.1^{+0.7}_{-0.3}$ |
| **$\phi_1$** | $4^{+9}_{-9}$ | $89^{+9}_{-11}$ | $-20^{+12}_{-10}$ |
| **$v_{r1}$** | $-390^{+103}_{-154}$ | $-318^{+38}_{-27}$ | $-956^{+93}_{-131}$ |
| **$\phi_2$** | 0 | 0 | $-54^{+4}_{-10}$ |
| **$v_{r2}$** | 0 | 0 | $214^{+67}_{-42}$ |
| **$\phi_3$** | 0 | 0 | $54^{+6}_{-2}$ |
| **$v_{r3}$** | 0 | 0 | $1050^{+61}_{-133}$ |
| **$f_a$** | 0 | 0 | $0.06^{+0.07}_{-0.10}$ |
| **<dM/dt>** | $-24^{+6}_{-11}$ | $-39^{+6}_{-4}$ | $-26^{+4}_{-7}$ |
| **Log $M_b$** | $10.68^{+0.07}_{-0.04}$ | $11.00^{+0.07}_{-0.05}$ | $11.00^{+0.30}_{-0.02}$ |
| **Log $j_b$** | $3.82^{+0.16}_{-0.03}$ | $3.60^{+0.07}_{-0.09}$ | $4.04^{+0.09}_{-0.02}$ |
| **$\lambda_b$** | $0.26^{+0.08}_{-0.06}$ | $0.70^{+0.04}_{-0.03}$ | $0.19^{+0.11}_{-0.01}$ |

**Supplementary Table 8 – CSO38B Continuum Object Parameters and Errors**

| Object | BX173 | MD23 |
|---|---|---|
| **Δv** | -170 km/s | -490 km/s |
| **E(B-V)** | $0.18\ [-0.01, +0.01]$ | $0.22\ [-0.01, +0.01]$ |
| **Log (SFR)** | 1.04 [-0.041, +0.01] | 1.83 [-0.064, +0.047] |
| **Log ($M_*$)** | $10.35\ [-0.015, +0.021]$ | $9.63\ [-0.028, +0.040]$ |
| **Log (Age/yr)** | $9.30$ $[\max\ allowed\ by\ modelling]$ | $7.80\ [-0.0, +0.10]$ |

**Supplementary Notes**

## Velocity Map Errors

The examples given in Supplementary Figure 7 show that because these nebulae are quite bright, the velocity errors due to Poisson fluctuations are low even in the lowest flux areas for both objects. The errors peak at ~50 km/s in small regions with a large velocity gradient and lower flux. As discussed in Methods, in order to derive conservative chi-square and Akaike Information Criteria (AIC) values we use the minimum rms velocity residual in place of the statistical error in calculating chi-square. Since the rms velocity residual is >72 km/s in both objects the statistical velocity centroid error has neglible effect. The implication is that all error limits on model parameters are conservative. In particular, the error limits on the MFI components and radial mass flux are conservative.

Because the signal-to-noise ratio is high, demonstrated by our investigation of the impact of the SNR threshold, variation in the details of the smoothing algorithm will not impact the results in a significant way. In an earlier paper [1] we gave extensive tests of the algorithm for much lower SNR data, showing that it is robust.

We have also estimated the contribution of sky subtraction error. This contribution depends on the smoothness of the sky spectrum at the systemic Lyα wavelength. A perfectly flat, smooth sky spectrum does not perturb the velocity centroid even with a 100% subtraction error. Assuming that the line flux has a peak equal to the sky background, and that there is a 100% subtraction error, the velocity error for CSO38B is 13 km/s (mostly from a slow continuum slope) and UM287 is 4 km/s. The faintest area of each nebula exceeds 0.1 times sky, and a sky subtraction error of <1% is estimated from the lack of residual sky features and previous experience. Thus the typical error contribution would be 1.3 km/s for CSO38B and 0.4 km/s for UM287. Even a sky subtraction error of 10% would have no effect on the results.

### Previously proposed criteria for MFI Proto-galaxies

1. Higher intensity symmetrically located around 1D velocity center and relatively uniform intensity with clear intensity break at edges;
2. Near constant slope velocity gradient (1D) consistent with an NFW halo;
3. 2D velocity and intensity distribution consistent with a disk in an NFW halo with minimal residuals;
4. Evidence for one or more filaments with low velocity gradient and possibly lower-velocity dispersion;
5. Abrupt kinematic transition at disk edge ( versus continuous acceleration expected from single-filament dragfree infall) either in mean velocity or velocity dispersion or both consistent with filament/ disk interface
6. Star formation at center of disk co-located with a possible intensity/ gas deficit; and
7. Kinematics consistent with radial and spiral inflow.

### Description of Continuum Objects near CS038B

Supplementary Table 8 gives derived spectral energy distribution fitting parameters and errors of two of these objects. Object BX173 is a star forming galaxy with redshift z=2.6499, close to the redshift of the QSO and the nebula (z=2.652, giving Δv=-170 km/s). BX173 falls near the intensity peak, center of light, and kinematic center of the nebula. From broad-band photometry BX173 has an inferred star formation rate of $17\ M_{\odot} yr^{-1}$, a stellar mass of $3.5 \times 10^{10}\ M_{\odot}$, and a corresponding mass doubling time of 2000 Myr. This suggests that this galaxy is well established and has been forming stars for more than 1 Gyr. The inferred nebula mass is |$\sim 8 \times 10^{10}\ M_{\odot}$ (Supplementary Table 5). Object MD23 lies at the southern tip of the nebula. It has a redshift of z=2.646, giving Δv=-490 km/s, somewhat outside the kinematic range of the nebula. From photometry has a star formation rate of $27\ M_{\odot} yr^{-1}$, a stellar mass of $7 \times 10^{9}\ M_{\odot}$, and a corresponding mass doubling time of 260 Myr. Two other continuum objects are detected in the eastern part of the nebula. These do not have redshifts or photometry and SED fitting. Neither overlaps the local maximum in the nebular intensity. If they are at the nebular redshift, and based on their rest-frame UV fluxes, their star formation rates would be estimated to be $56\ M_{\odot} yr^{-1}$ and $17\ M_{\odot} yr^{-1}$.

We note that all the objects detected have inferred star formation rates that are consistent with the radial mass flux estimated for Model 3a(i). Furthermore, the direction of gas flow is such to direct gas to BX173 and MD23 (Fig. 4), as seen in projection. The simulated galaxy VELA07 has a star forming galaxy at the center with a similar stellar mass, star formation rate, and age as that estimated for BX173. It also has several other stellar objects with lower stellar mass and star formation rates but similar ages.

**Description of Continuum Sources near UM287**

Of the sources near UM287, Source C is the brightest and shows CIV1549 and HeII1640 as well as Hα. This source shows the largest redshift and is not perfectly fit by the MFI model (see Supplementary Fig. 6bcd, with d showing a deviation of ~100 km/s at source C). If this local source has an outflow it could produce additional redshift due to the usual blueshifted absorption. There is however an extended region near source C which is has a redshift consistent with the MFI model. Source C could increase the illumination of the nebula around it over that of the QSO. Sources D and E are close to the light and kinematic center of UM287. The rest frame UV continuum star formation rate we infer for Source D and E (>22 $M_{\odot}/yr$), is consistent with the overall mass influx from the Model 3 and 4 fits. There are potentially other, fainter extended continuum sources. All of these are consistent with clumpy star formation regions that might be expected in a forming, non-equilibrium protogalaxy.

**Impact of Selecting Objects with Neighboring QSO**

It should be noted that we should use caution when applying descriptions of low redshift phenomena at high redshift. At the redshifts of these observations, simulations predict that galaxies undergoing cold spiral inflow exhibit continuous smooth and clumpy accretion (i.e., continuous “merging”). They tend to be in overdense environments with correspondingly high probability for nearby galaxies. Both the simulated galaxy and neighboring objects grow massive bulges and super massive black holes which will exhibit QSO-mode accretion occasionally during their growth. The projected separation of the observed objects and their nearby illuminating QSO is sufficient (greater than 160 kpc for CSO38B and greater than 100 kpc including the line of sight separation [2] for UM287) that tidal effects on the velocity fields (the main subject of this work) can be ignored (see discussion above). The main impact of the nearby QSO-mode accretion is to illuminate the neighboring forming galaxy we are observing, allowing us to make a sensitive, high-resolution map of the gas velocity field. The physical trigger of QSO-mode accretion is still controversial, as is the typical duration and duty cycle. Thus while it was not possible to select, in this initial study, simulation objects for zoom-in that exactly reproduce the conditions and demographics appropriate for our observed objects (e.g., with a nearby neighbor undergoing QSO-mode supermassive black hole accretion), the gas flows measured in the observed objects can be reasonably assumed to be representative of those in the simulations, as we have noted before [2,3].

For an alternative scenario based on recent MUSE observations of HeII detected in some regions of the UM287 nebula see ref[4].